\documentclass[journal]{IEEEtran}

\usepackage{cite}

\ifCLASSINFOpdf
  \usepackage[pdftex]{graphicx}
\else
\fi

\usepackage{amsmath, amssymb}
\usepackage{units}
\usepackage{bm}
\usepackage[ruled]{algorithm2e}

\usepackage{array}

\usepackage[caption=false, font=footnotesize]{subfig}
\usepackage{url}

\usepackage{xcolor}

\newcolumntype{L}[1]{>{\raggedright\let\newline\\\arraybackslash\hspace{0pt}}m{#1}}
\newcolumntype{C}[1]{>{\centering\let\newline\\\arraybackslash\hspace{0pt}}m{#1}}
\newcolumntype{R}[1]{>{\raggedleft\let\newline\\\arraybackslash\hspace{0pt}}m{#1}}

\newcommand{\svec}{{\bf{s}}}
\newcommand{\ivec}{{\bf{i}}}
\newcommand{\vvec}{{\bf{v}}}
\newcommand{\evec}{{\bf{e}}}
\newcommand{\cvec}{{\bf{c}}}
\newcommand{\uvec}{{\bf{u}}}
\newcommand{\hvec}{{\bf{h}}}
\newcommand{\pvec}{{\bf{p}}}
\newcommand{\Ymat}{{\bf{Y}}}
\newcommand{\Umat}{{\bf{U}}}
\newcommand{\C}{\mathbb{C}}
\newcommand{\R}{\mathbb{R}}
\newcommand{\Imat}{{\bf{I}}}
\newcommand{\Lmat}{{\bf{L}}}
\newcommand{\Mmat}{{\bf{M}}}
\newcommand{\Vmat}{{\bf{V}}}
\newcommand{\zerovec}{{\bf{0}}}
\newcommand{\onevec}{{\bf{1}}}
\def\psivec{{\mbox{\boldmath $\psi$}}}
\newcommand{\varPhivec}{{\bm{\varphi}}}
\newcommand{\Lambdamat}{{\bf{\Lambda}}}
\newcommand{\define}{\stackrel{\triangle}{=}}

\begin{document}
\bstctlcite{IEEEexample:BSTcontrol}

\title{Detection of False Data Injection Attacks in Smart Grids based on Graph Signal Processing}

\author{Elisabeth~Drayer,~\IEEEmembership{Member,~IEEE},
        and~Tirza~Routtenberg,~\IEEEmembership{Senior Member,~IEEE}
\thanks{E. Drayer and T. Routtenberg are with the Department of Electrical and Computer Engineering, Ben-Gurion University of the Negev, Beer Sheva,
Israel e-mail: drayer@post.bgu.ac.il,~tirzar@bgu.ac.il. }
\thanks{This research was supported by the ISRAEL SCIENCE FOUNDATION (ISF), Grant No. 1173/16, by the Kreitman School of Advanced Graduate Studies, and by the BGU Cyber Security Research Center. 
} %
}

\maketitle

\begin{abstract}
The smart grid combines the classical power system with information technology, leading to a cyber-physical system. In such an environment the malicious injection of data has the potential to cause severe consequences. Classical residual-based methods for bad data detection are unable to detect well designed false data injection (FDI) attacks. Moreover, most work on FDI attack detection is based on the linearized DC model of the power system and fails to detect attacks based on the AC model. The aim of this paper is to address these problems by using the graph structure of the grid and the AC power flow model. We derive an attack detection method that is able to detect previously undetectable FDI attacks. This method is based on concepts originating from graph signal processing (GSP). The proposed detection scheme calculates the graph Fourier transform of an estimated grid state and filters the graph's high-frequency components. By comparing the maximum norm of this outcome with a threshold we can detect the presence of FDI attacks. Case studies on the IEEE 14-bus system demonstrate that the proposed method is able to detect a wide range of previously undetectable attacks, both on angles and on magnitudes of the voltages.
\end{abstract}


%

\section*{Abbreviations}
\addcontentsline{toc}{section}{Nomenclature}
\begin{IEEEdescription}[\IEEEusemathlabelsep\IEEEsetlabelwidth{SCADA}]
\item[AC]	Alternating current
\item[DC]	Direct current
\item[DSP]	Digital signal processing
\item[FDI]	False data injection
\item[GFT]	Graph Fourier transform
\item[GHPF]	Graph high-pass filter
\item[GSP]	Graph signal processing
\item[ICT]	Information and communication technology
\item[PMU]	Phasor measurement unit
\item[PSSE]	Power system state estimation
\item[SCADA]	Supervisory control and data acquisition
\end{IEEEdescription}

\section{Introduction}
\label{sec:Introduction}
\IEEEPARstart{T}{he} vulnerability of critical infrastructures to cyber attacks is a major threat to the stability and safety of our society \cite{nato2016CriticalInfrastructureProtection}. One of the most critical infrastructures is the power system, as almost all other infrastructures depend on it \cite{lewis2006CIProtection}. In particular, smart grids that combine classical power system components with advanced information and communication technology (ICT) to build a cyber-physical system are considered to be particularly vulnerable, due to (among other reasons) dedicated attacks \cite{buldyrev2010CatastrophicCascade, teixeira2015SecureControlSystems, smith2016GoingBeyondCybersecurity, soltan2018CyberPhysicalAttack}.

Research on power system operation is used to develop methods to cope with physical faults or disturbances in power systems, such as outages \cite{drayer2018DistributedSelfHealing}, off-nominal frequencies \cite{routtenberg2013JointFrequency}, and voltage imbalances \cite{routtenberg2017PUMBasedDetection, routtenberg2018CentralizedIdentification}. However, the development of methods to handle and mitigate the effects of cyber attacks in power systems is a rather new field, and traditional security measures originating from the ICT domain are considered to be insufficient against this growing danger \cite{jin2018PowerGridACStateEstimation}. 

In this paper, we consider the problem of detection of false data injection (FDI) attacks, in which it is assumed that attackers can compromise the measurements of the grid. Such FDI attacks aim to affect the power system state estimation (PSSE). The state of a power system is generally defined as the voltage values on all the buses of the system. The PSSE is part of the regular control routines that are hosted in the supervisory control and data acquisition (SCADA) system of the control center of the grid operator. It provides the input for multiple monitoring and control purposes, including security assessment, load forecasting, dispatching, reliability analysis, and economic considerations \cite{abur2004PowerSystemStateEstimation, giannakis2013MonitoringOptimization}. In particular, the modern power system, which is often operated near its operational limits for economic reasons, cannot be operated without a reliable PSSE. Thus, the impact of FDI attacks could be manifold and range from economic consequences, through overloading and physical damage, to serious human hazard \cite{liang2017ReviewFalseDataInjection, kim2015AgainstDataAttacks, liang2016VulnerabilityAnalysisFalseData}. The PSSE is usually equipped with methods to detect random false data and faults that are based on residuals (see, e.g. \cite{handschin1975BadData, monticelli1983ReliableBadData} and Ch. 5 in \cite{abur2004PowerSystemStateEstimation}). However, it was shown in \cite{liu2011FalseDataInjection} that if the attacker has sufficient knowledge of the system topology, a well-designed attack can pass the residual-based bad data detector and perturb the PSSE to any desired level. This knowledge about the system topology can be gained by analyzing measurements \cite{li2013BlindToplogyIdentification, anwar2016EstimationGridToplogy, gera2017BlindEstimation, cavraro2018GraphAlgorithmToplogyIndenfication, grotas2018PowerSystemsTopology}. Such types of attacks are called \emph{undetectable attacks} \cite{liu2011FalseDataInjection} or \emph{stealth attacks} \cite{vukovic2012NetworkAwareMitigation, wang2017EffectsSwitchingNetworkTopologies}. Extensions of the classical residual-based methods have been suggested, e.g. the work in \cite{kosut2011MaliciousDataAttacks} that compares the norms of attacked states against those of unattacked ones, or the work of \cite{huang2016RealTimeDetection} that proposes an adapted cumulative sum method, but these, too, cannot overcome the model's inherent limitations in its detection capabilities of such attacks.

Seeking to derive ways of rendering the smart grid robust to FDI attacks, the authors of \cite{liang2017ReviewFalseDataInjection} review major types of defense strategies, such as data authentication \cite{kim2015AgainstDataAttacks}, and the inclusion of time synchronized phasor measurement units (PMUs) \cite{jiang2017DefenseMechanisms, ashok2018OnlineDetectionStealthy}. In these works, the measurements originating from PMUs are considered to be protected against attacks, and thus can be used as a trustworthy reference. However, these methods would require the placement of additional expensive hardware into the grid infrastructure.

Approaches based on data analytics, on the other hand, try to detect the presence of bad data or FDI attacks by analyzing the measurements and the outcome of the PSSE, while not requiring new hardware. In \cite{liu2014DetectingFalseData} the authors propose a method to detect previously unobservable attacks, but their approach is limited to sparse attacks where the measurement devices are only compromised temporarily. In \cite{jiang2017DefenseMechanisms} the residual evaluation is done between two consecutive time steps, thus facilitating the detection of unobsorvable attacks. However, this approach is easily corrupted, especially if the attack is launched with small gradual changes. Other detection methods for undetectable FDI attacks require reliable load forecasts \cite{ashok2018OnlineDetectionStealthy}, or explore machine learning concepts \cite{esmalifalak2017DetectionStealthy}. The main problem with machine learning approaches, such as in \cite{esmalifalak2017DetectionStealthy}, is that they need a large data set of historic grid states that also contains attacked states. This can be very difficult to provide. Finally, in \cite{xu2018AchievingEfficientDetection} and \cite{leirzapf2017NetworkForensicAnalysis}, signals originating from hardware components and from the process level, respectively, are used to detect unobservable FDI attacks, but these solutions are very specific to the hardware and protocols used in the power system. 

Most of the aforementioned works, notably \cite{kosut2011MaliciousDataAttacks, huang2016RealTimeDetection, liu2014DetectingFalseData, jiang2017DefenseMechanisms, yu2015FalseDataPCA, esmalifalak2017DetectionStealthy, ashok2018OnlineDetectionStealthy}, rely on the linearized DC model to describe the electrical behavior of the power system. However, since the DC model is only a simplification of the more realistic and more accurate AC model, this is a significant shortcoming with regard to the vulnerability analysis \cite{jin2018PowerGridACStateEstimation}. Since the construction of an undetectable attack based on the AC model requires that the attacker has far more knowledge and can solve non-convex optimization problems \cite{kosut2011MaliciousDataAttacks, jin2018PowerGridACStateEstimation}, detection methods for AC model-based FDI attacks have not been extensively studied yet (see, e.g. \cite{chaojun2015DetectinFalseData}). Nevertheless, several contributions that show how an attacker could construct undetectable attacks based on the AC model do exist (see, e.g. \cite{kosut2011MaliciousDataAttacks, hug2012VulnerabilityAssessmentAC, liu2017FalseDataAttackAC, jin2018PowerGridACStateEstimation}). 

In recent years, the research field of graph signal processing (GSP) has emerged. The key idea of GSP is to extend the concepts of digital signal processing (DSP) to data connected on graphs \cite{shuman2013EmergingFieldGSP, sandryhaila2014DSPonGraph, teke2017ExtendingClassicalMultirate, marques2017StationaryGraphProcesses, ortega2018GraphSignalProcessing}. This enables the definition of classic signal processing concepts, such as filtering, sampling, and modulation, for signals related to an underlying graph structure \cite{shuman2013EmergingFieldGSP}. Previously, GSP has been applied to various fields, such as sensor networks, biological networks, image processing, machine learning, and data science  \cite{ortega2018GraphSignalProcessing, huang2018GSPBrain}. However, GSP has rarely been applied in the context of power systems (see, e.g. \cite{he2018NonIntrusiveLoadDisaggregation}, where it is used to disaggregate the total load of one power measurement).

In this paper, we consider the detection of FDI attacks in power systems based on the output of the PSSE. We use the concept of the graph Fourier transform (GFT), originating from GSP and inspired by the work presented in \cite{sandryhaila2014DSPonGraph} that detects anomalies in a network of temperature sensors. By relying on the electrical properties of the power system, which can be interpreted as an undirected graph, we design a new detection method that is not limited by fundamental unobservabilities that originate from the simple linear equations of the system and, thus, can detect unobservable attacks.

The contribution of this paper is threefold. First, we propose a statistical model of the output of general PSSE, which can be based on any type of measurement, including measurements form smart meters or PMUs. Second, this approach relies on the AC power system model and, thus, facilitates the detection of attacks on both voltage angle and magnitude. Third, we derive a new method to detect previously undetectable FDI attacks in power systems. The proposed method is based on filtering the graph Fourier transformed estimated grid state. This filter is designed as a high-pass filter in the sense of graph frequencies. Then, by comparing the maximum norm of the filtered signal with a threshold, the presence of FDI attacks is discovered. In addition, the influence of the graph smoothness on the detection of FDI attacks is investigated. Finally, we conduct numerical simulations on the IEEE 14-bus test case that demonstrate that the proposed method is able to detect previously undetectable attacks. A preliminary version of this approach, which is limited to the DC model and to the detection of voltage angles and without the derivations of the cutoff frequency of the filter and the detection threshold has been published in \cite{drayer2018BadDataGSP}.

The remainder of the paper is organized as follows. In Section~\ref{sec:psModel} we introduce the mathematical modeling of the power system and the FDI attacks. The use of GSP concepts for FDI attack detection is presented in Section~\ref{sec:AnomalyDetection}. Several case studies in Section~\ref{sec:CaseStudy} show the successful implementation of the proposed method. In Section~\ref{sec:Remarks} the results of the case studies are examined in detail and the use of the proposed method for the DC power flow model is outlined. The paper ends with conclusions in Section~\ref{sec:Conclusion}. 

\section{Power System and Attack Modeling}
\label{sec:psModel}
In this section, we first model the power system as an undirected graph in Subsection~\ref{sec:GraphPowerSystem}. In Subsection~\ref{sec:ACmodel} we present the AC power flow model used in this paper. The considered FDI attack detection is presented in Subsection~\ref{sec:Hypothesis} in the form of a hypothesis testing problem.

\subsection{Graph Representation of Power System}
\label{sec:GraphPowerSystem}
The power system can be represented by an undirected graph, $\mathcal{G}=\left(X,E\right)$, where $X = \{1,2,\dots, M \}$ is a set of $M$ nodes that represent the buses with connected loads or generators, and $E = \{\left(e_{k,l}\right)\}$ is a set of edges connecting the buses, in which $e_{k,l}$ represents the transmission line between bus $k$ and bus $l$, for any $k,l \in X$ where there is a transmission line between those buses. As the electric characteristic of a line is independent of the direction of the current that goes through it, the graph is assumed to be undirected \cite{giannakis2013MonitoringOptimization}. Such finite graph structures are described by a weighted Laplacian matrix, $\Ymat$, with the following $(k,l)$-th element 
\begin{equation}
\label{equ:YForm}
\displaystyle \Ymat_{k,l} = \begin{cases}
	\displaystyle \sum_{E_k} y_{k,m},& k = l \\
  \displaystyle  - y_{k,l},& k \neq l \text{ and there exists } e_{k,l} \\
     \displaystyle 0,& \text{otherwise}
\end{cases},
\end{equation}
where $E_k = \{e_{k,m} \in E, m = 1,2,\dots, M\}$, and $y_{k,l}$ is the weight assigned to the connection between node $k$ and node $l$. In the power system context, this weight represents the electrical admittance, $y_{k,l}$, of the particular transmission line, and $\Ymat$ is also called the admittance matrix. Thus, the matrix $\Ymat$ depends on the topology of the grid, as well as on the admittance values of the lines. With regard to the modeling of the electrical behavior of the power system via the power flow equations, one can distinguish between the approximated linearized DC power flow model and the more accurate non-linear AC model \cite{wood1984PowerGenerationOperation}. In the following, the AC model is considered. A summary of how to use the proposed method on the DC model is given in Subsection~\ref{sec:DCmodel}. 

\begin{figure}[t]
 \centering
 \includegraphics[width=3.5 in]{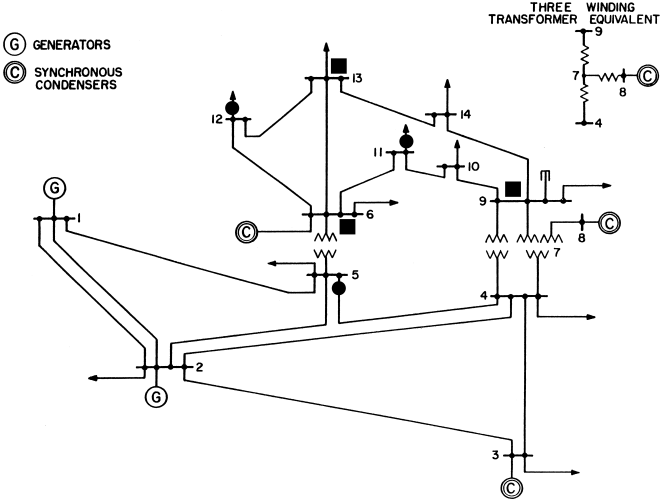}
 \caption{IEEE 14-bus system based on \cite{scheme14bus2019}. Circles represent sample power measurements and rectangles represent sample voltage measurements.}
\label{fig:IEEE14realScheme}
\end{figure}

To visualize this step, Fig.~\ref{fig:IEEE14realScheme} gives the one-line diagram of the IEEE 14 bus system, which is a well-known test grid for power system applications. It is used as the basis for our case studies in Section~\ref{sec:CaseStudy}. Figs.~3(a) and 3(b) give the graph representation of this grid.

\subsection{AC Model}
\label{sec:ACmodel}
The goal of the AC power flow analysis is to compute the complex voltages at each bus in steady-state conditions \cite{wood1984PowerGenerationOperation}. These voltage values are considered to be the system state of the power system. The AC power flow equations can be written in matrix form as follows:
\begin{equation}
\label{equ:PowerFlowAC}
\ivec = \Ymat \vvec,
\end{equation}
where $\ivec = \left[i_1,i_2, \dots, i_{M}\right]^T \in \C^M$ is the vector of currents at the $M$ buses, $\vvec = \left[v_1, v_2, \dots, v_{M}\right]^T \in \C^M$ is the vector of bus voltages, and $\Ymat \in \C^{M \times M}$ is the admittance matrix, as defined in (\ref{equ:YForm}). The vector of the complex bus currents, $\ivec$, is related to the apparent power (either load or generator) attached to the buses, $\svec= \left[s_1, s_2, \dots, s_{M}\right]^T \in \C^M$, according to 
\begin{equation}
\label{equ:PQNode}
\svec = {\text{diag}}(\vvec) \ivec^*,
\end{equation}
where $\ivec^*$ is the complex conjugate of the vector of bus currents. Inserting (\ref{equ:PowerFlowAC}) into (\ref{equ:PQNode}) results in 
\begin{equation}
\label{equ:PowerFlowACII}
\svec = {\text{diag}}(\vvec) \left(\Ymat \vvec \right)^*.
\end{equation}
This reveals that the AC power flow model is non-linear and non-convex, and, thus, it is usually solved by numerical methods.

Generally, the complex voltage at the $m$-th bus in polar form is given by $v_m = V_m e^{j \varphi_m} \; \forall \; m=1,2,\dots,M$, where $V_m$ is the voltage magnitude in per unit (p.u.) and $\varphi_m$ is the voltage angle or phase. Neglecting the phase shift and off nominal turns ratio (tap) of transformers, as well as the capacitive shunt reactances of lines, the admittance, $y_{k,l}$, of each line in the electric grid is described by
\begin{equation}
\label{equ:admittanceAC}
y_{k,l} = \frac{1}{r_{k,l}+j x_{k,l}},~\forall e_{k,l} \in E,
\end{equation}
where $r_{k,l}$ and $x_{k,l}$ denote the electrical resistance and reactance, respectively. By substituting (\ref{equ:admittanceAC}) in (\ref{equ:YForm}) we obtain the admittance matrix for the AC model, $\Ymat$, which can be interpreted as a complex, non-Hermitian Laplacian matrix. Thus, under the AC model, the eigenvalues of $\Ymat$ generally have complex values that do not have a partial order \cite{rudin1987RealComplexAnalysis}. The definition of ordering of the eigenvalues in this case is a theoretical open question in GSP \cite{shuman2013EmergingFieldGSP, sandryhaila2014DSPonGraph, ortega2018GraphSignalProcessing}. To overcome the problem of complex ordering in this paper, we propose to decompose the complex matrix $\Ymat$ as follows: 
\begin{equation}
\label{equ:Ydecompose}
\Ymat = \Ymat^\text{R} + j \Ymat^\text{J},
\end{equation}
where $\Ymat^\text{R} \in \R^{M \times M}$ and $\Ymat^\text{J} \in \R^{M \times M}$ are the real and imaginary parts of $\Ymat$, respectively. By substituting (\ref{equ:Ydecompose}) into (\ref{equ:PowerFlowAC}) we obtain
\begin{equation}
\label{equ:ACdecompose}
\ivec = \Ymat^\text{R} \vvec^\text{R} + j \Ymat^\text{J} \vvec^\text{R} + j \Ymat^\text{R} \vvec^\text{J} - \Ymat^\text{J} \vvec^\text{J},
\end{equation}
where $\vvec^\text{R}$ and $\vvec^\text{J}$ are the real and imaginary parts of $\vvec$, respectively, such that $\vvec=\vvec^\text{R}+j\vvec^\text{J}$. The matrices $\Ymat^\text{R}$ and $ - \Ymat^\text{J}$ are both real Laplacian matrices. 

\subsection{Hypothesis Testing}
\label{sec:Hypothesis}
\begin{figure}[t]
 \centering
 \includegraphics[width=3.5 in]{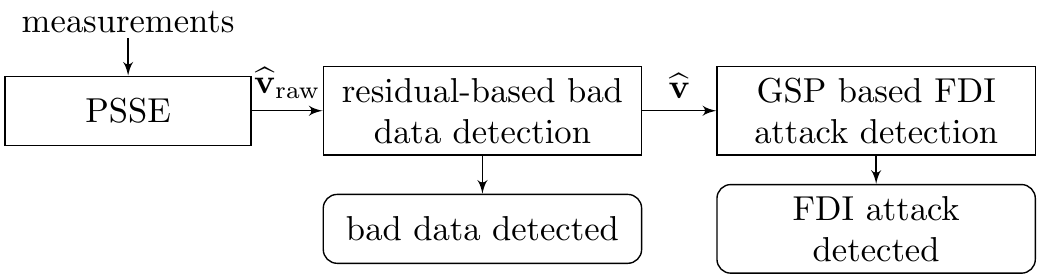}
 \caption{Conceptional PSSE and bad data detection routine supplemented by the proposed FDI attack detection method.}
 \label{fig:steps}
\end{figure}

In the following, it is assumed that an attacker launches an undetectable FDI attack by tampering with some of the measurements of the power system. We assume that this attack can be on any type of measurement that is used as an input to the PSSE, such as classic measurements available in the SCADA, such as line power and current flows, power injection at the buses and bus voltage magnitudes, but also bus voltage angles originating from PMUs, and power consumption originating from smart meters. Thus, an advantage of the considered model is that it is not limited to specific measurements and can be used for both smart and traditional systems. The PSSE uses the measurements and calculates the system state, $\widehat{\vvec}$, i.e. the voltage angle and magnitude for every bus in the power system. As the attack is assumed to be undetectable for classic approaches, it bypasses the PSSE and the residual-based bad data detection. Therefore, in this work we consider that the output of the PSSE is given by
\begin{equation}
\label{equ:xbad}
\widehat{\vvec} = \vvec + \evec + \cvec,
\end{equation}
where $\vvec$ is the true unknown value of the bus voltage vector, $\evec$ is the error of the PSSE, which is assumed to be a zero-mean Gaussian noise vector with known variance, $\sigma_\text{e}^2$, and $\cvec \in \C^{M}$ is an arbitrary vector defining the impact of the attack. If $\cvec=\zerovec$, there is no attack. It should be emphasized that while FDI attacks are usually considered to be sparse attacks \cite{kosut2011MaliciousDataAttacks, liu2014DetectingFalseData, hao2015SparseMaliciousFDI}, in the sense that attacks are launched only at a few buses, here, the attack vector $\cvec$ is not necessarily sparse, since it represents the influence of the attack on the PSSE. 
The PSSE output in (\ref{equ:xbad}) is very general and can be obtained using any existing PSSE method. That is, in this paper we take the PSSE for granted and use its result to detect the attacks. Fig.~\ref{fig:steps} illustrates the standard PSSE and bad data detection routine with an additional FDI attack detection method. 
 
The task of detecting an FDI attack based on the PSSE output is equivalent to making the decision if $\cvec=\zerovec$ in (\ref{equ:xbad}). In hypothesis testing formulation, we consider the following problem: 
\begin{equation}
\label{equ:hypothesis}
\left\{  \begin{aligned}
\mathcal{H}_0 &: \cvec = 0 \\
\mathcal{H}_1 &: \cvec \neq 0,   
  \end{aligned} \right.
\end{equation}
where $\mathcal{H}_0$ represents the hypothesis of no attack and $\mathcal{H}_1$ represents the hypothesis of an FDI attack. Since we do not know the true value of $\vvec$ in (\ref{equ:xbad}), we cannot distinguish between $\vvec$ and $\vvec +\cvec$ and, thus, this type of attack cannot be detected by likelihood ratio tests. In this paper, we rely on the inherent graph structure of the electrical properties of the grid to gain additional insights into the situation and to overcome this limitation.

\section{FDI Attack Detection}
\label{sec:AnomalyDetection}
In this section we develop a novel method for the detection of unobservable FDI attacks.
As this approach requires real Laplacian matrices as inputs, each of the terms in (\ref{equ:ACdecompose}) needs to be investigated independently. For the sake of simplicity of presentation, all the following derivations are presented for the first term in (\ref{equ:ACdecompose}), taking the real part of the Laplacian matrix, $\Ymat^\text{R}$, and the real part of the voltage vector, $\vvec^\text{R}$. For the other three terms the approach is analogous, and comprises the following steps: After verifying the small total variation of the graph (see Subsection~\ref{sec:Smoothness}) and the eigendecomposition, the first step is to calculate the GFT, as described in Subsection~\ref{sec:GFT}. The second step is to design the graph high-pass filter (GHPF), as described in Subsection~\ref{sec:GHPF}. The third step is the thresholding and detection, as described in Subsection~\ref{sec:Detection}. 

\subsection{Smooth Graphs}
\label{sec:Smoothness}
The smoothness of a signal defined on graph vertices can be measured by the total variation \cite{shuman2013EmergingFieldGSP, zhu2012ApproximatingSignalsOnGraphs}:
\begin{equation}
\label{equ:smoothnes}
S \left( \widehat{\vvec}^\text{R} \right)= \frac{1}{2} \sum_{k \in X} S_{k} \left( \widehat{\vvec}^\text{R} \right), 
\end{equation}
where 
\begin{equation}
\label{equ:smoothnesLocal}
S_{k} \left( \widehat{\vvec}^\text{R} \right) =  \sum_{l \in \mathcal{N}_k} \Ymat^\text{R}_{k,l} \left(\widehat{v}^\text{R}_{k} - \widehat{v}^\text{R}_{l} \right)^2
\end{equation}
is the local variation at vertex $k$. It can be seen that $S_{k}$ is a function of all differences between the state of bus $k$ and the states of all its neighboring buses, $l \in \mathcal{N}_k$, normalized by the particular entry of the admittance matrix, $\Ymat_{k,l}$. The smaller the total variation is, the smoother is the signal defined on the graph. A sufficient smoothness of the undisturbed signal is a necessary condition for detection after high-pass filtering. The total variation can also be an indicator of the presence of bad data or FDI attacks, as such disturbances tend to increase the total variation defined in (\ref{equ:smoothnes}). As a visualization, in Figs.~3(a) and 3(b) the grid of the IEEE 14-bus system is represented as a graph. The color of the node indicates the variation per node, $S_{k}$, as defined in (\ref{equ:smoothnesLocal}), for an unattacked (Fig.~3(a)) and an attacked (Fig.~3(b)) grid state. It can be seen that the unattacked grid is much smoother than the attacked grid. 

 \begin{figure}[t!]
 \centering
     \subfloat[ \label{fig:IEEE14graphNoAttack}]{
      \includegraphics[width=2.6in]{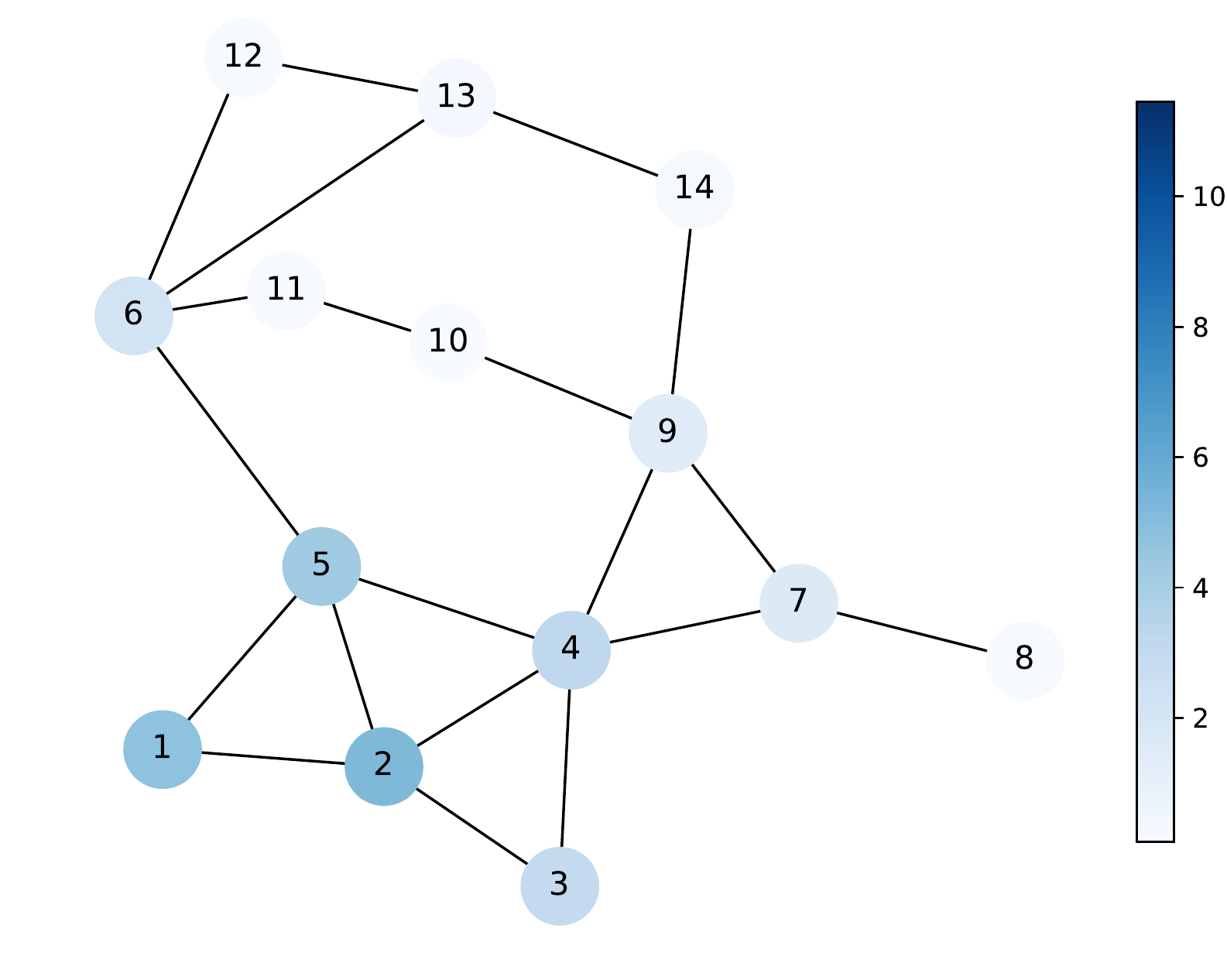}
      }
      \hfill
      \subfloat[\label{fig:IEEE14graphAttack}]{
        \includegraphics[width=2.6in]{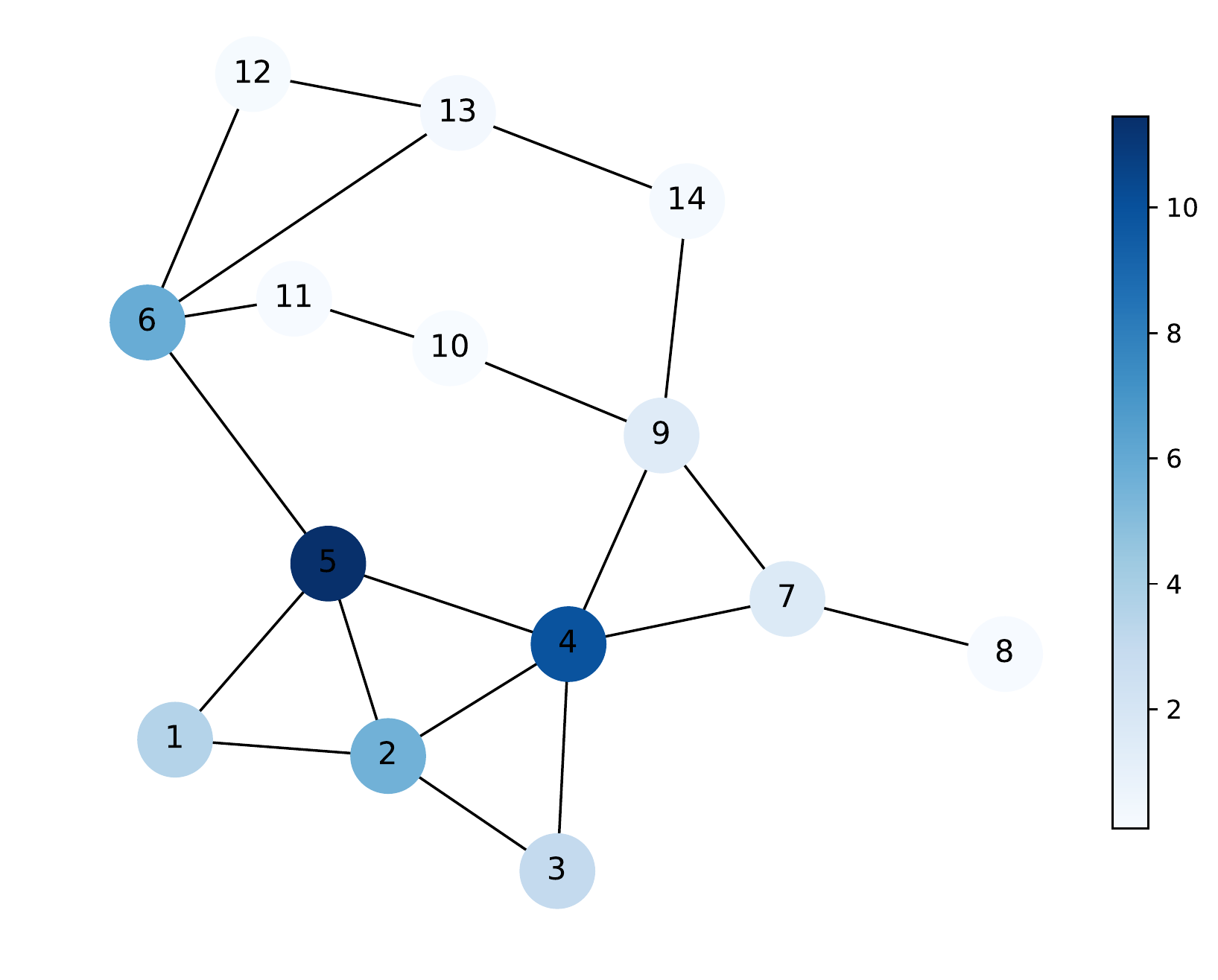}
	} 
	\caption{Graph representation of IEEE 14-bus system from Fig.~\ref{fig:IEEE14realScheme}. The node color represents the local variation as defined in (\ref{equ:smoothnesLocal}). In Fig.~3(a) the grid is not under attack, while in Fig.~3(b) node 5 is attacked.}
 \end{figure}

\subsection{Graph Fourier Transform (GFT)}
\label{sec:GFT}
The eigendecomposition of the real, symmetric positive-semidefinite matrix $\Ymat^\text{R}$ defined in Subsection~\ref{sec:ACmodel} results in a set of ordered real eigenvalues $\lambda_{1},\lambda_{2}, \dots, \lambda_{M}$ that satisfy
\begin{equation}
\label{equ:lambdaOrder}
0 = \lambda_1 < \lambda_2 \leq  \dots \leq \lambda_{M},
\end{equation}
and a set of orthonormal eigenvectors $\Umat = \left[ \uvec_1,\uvec_2, \dots,\uvec_M\right]$ that satisfy the following matrix decomposition:
\begin{equation}
\label{equ:decomposition}
\Ymat^\text{R} = \Umat^\text{R} \Lambdamat^\text{R} \left(\Umat^\text{R}\right)^T,
\end{equation}
where $\Lambdamat^\text{R}$ is a diagonal matrix with the eigenvalues from (\ref{equ:lambdaOrder}). In the context of GSP, the operation of (\ref{equ:decomposition}) is referred to as spectral decomposition of the matrix $\Ymat^\text{R}$, the eigenvectors are called the \emph{spectral components} and the eigenvalues can be interpreted as the \emph{graph frequencies} \cite{ortega2018GraphSignalProcessing}. The ordering of the real eigenvalues in (\ref{equ:lambdaOrder}) represents the ordering of the graph frequencies, ranging from low to high frequencies.  

Based on the decomposition in (\ref{equ:decomposition}), the GFT of the signal $\widehat{\vvec}^\text{R}$ is defined as
\begin{equation}
\label{equ:GFT}
\mathcal{F}_\text{R}\{\widehat{\vvec}^\text{R}\} =\left(\Umat^\text{R}\right)^T \widehat{\vvec}^\text{R},
\end{equation}
and the resulting signal, $\mathcal{F}_\text{R}\{\widehat{\vvec}^\text{R}\}$, has properties analogous to the Fourier transform of classic signals \cite{ortega2018GraphSignalProcessing, zhu2012ApproximatingSignalsOnGraphs}. In particular, for signals with a small total variation, i.e. high smoothness, the graph Fourier coefficients, i.e. the elements of the GFT vector defined in (\ref{equ:GFT}), decay with increasing graph frequency, having the largest contributions in the low-frequency components \cite{shuman2013EmergingFieldGSP}. The major assumption that enables the detection of FDI attacks based on the proposed approach is that for an attacked state the Fourier coefficients no longer decay, but they cause peaks in the high-frequency range of the Fourier coefficients.

\subsection{Filter Design: Graph High-Pass Filter (GHPF)}
\label{sec:GHPF}
Analogous to classical DSP theory, a graph filter is a system that takes a graph signal as an input, processes it, and produces another graph signal as an output. In the following, we consider polynomial graph filters that take the form \cite{sandryhaila2013DiscreteSignalProcessingGraphs}:
\begin{equation}
\label{equ:polyHwihtL}
h \left(\Ymat^\text{R} \right) = h_0\Imat + h_1\Ymat^\text{R} + \dots + h_L\left(\Ymat^\text{R}\right)^L,
\end{equation}
where $h$ is the transfer function of the filter and $L$ is the degree of the filter. By substituting the decomposition from (\ref{equ:decomposition}) in (\ref{equ:polyHwihtL}) we obtain:
\begin{equation}
\label{equ:fourierh}
 h \left(\Ymat^\text{R} \right) = \Umat^\text{R} h\left(\Lambdamat^\text{R}\right) \left(\Umat^\text{R}\right)^{T}, 
\end{equation}
where
\begin{equation}
\label{equ:polyH_Lambda}
 h \left(\lambda_i \right) = h_0+ h_1\lambda_i + \dots + h_L\lambda_i^L, \; i = 1,2,\dots, M.
\end{equation}
The output of the filter in (\ref{equ:fourierh}) is the signal:
\begin{equation}
\label{equ:filterOutput}
\widehat{\psivec}^\text{R} =  h \left(\Ymat^\text{R} \right) \widehat{\vvec}^\text{R}. 
\end{equation}
Therefore, according to (\ref{equ:GFT}), by multiplying (\ref{equ:filterOutput}) on the left by $\left(\Umat^\text{R}\right)^{T}$ the GFT of the output in (\ref{equ:filterOutput}) satisfies
\begin{equation}
\label{equ:filtering}
\mathcal{F}_\text{R}\{\widehat{\psivec}^\text{R}\} = \left(\Umat^\text{R}\right)^{T}  h \left(\Ymat^\text{R} \right) \widehat{\vvec}^\text{R}. 
\end{equation}
Now, by substituting (\ref{equ:GFT}) and (\ref{equ:fourierh}) in (\ref{equ:filtering}) and using the fact that for eigenvector matrices $\Umat^T \Umat=\Imat$, one obtains:
\begin{equation}
\label{equ:filteringGFT}
\mathcal{F}_\text{R}\{\widehat{\psivec}^\text{R}\} = h \left(\Lambdamat^\text{R}\right)\mathcal{F}_\text{R}\{\widehat{\vvec}^\text{R}\}. 
\end{equation}

The frequency response of a GHPF with the cutoff frequency $\lambda_\text{cut}$ is defined as follows:
\begin{equation}
\label{equ:frequencyResponse}
h\left( \lambda_i\right) =  \begin{cases}
    0,& \lambda_i \leq \lambda_\text{cut}\\
    1,& \lambda_i > \lambda_\text{cut}
\end{cases},\; i = 1, 2,\dots,M.
\end{equation}
The choice of $\lambda_\text{cut}$ is defined by (\ref{equ:cutOffError}) and further explained in Subsection~\ref{sec:Detection}. 

The construction of a filter with the frequency response of (\ref{equ:frequencyResponse}) in the form of (\ref{equ:polyH_Lambda}), which is polynomial of degree $L$, corresponds to solving a system of the following $M$ linear equations:
\begin{equation}
\label{equ:indicator}
h_0 + h_1\lambda + \dots + h_L\lambda^L = \onevec_{\lambda > \lambda_\text{cut}}, 
\end{equation}
where $\onevec_{A}$ is the indicator function of the event $A$. By solving these linear equations the $L+1$ unknown coefficients of the filter $\hvec$ can be obtained. In this work the filter is designed with $L+1=M$. This choice results in a single exact solution. The idea of this GHPF is to extract the high-frequency components of $\mathcal{F}_\text{R}\{\widehat{\vvec}^\text{R}\}$ that contain information about the presence of FDI attacks.

\subsection{Detection Method}
\label{sec:Detection}
To detect the presence of FDI attacks, we use the GHPF described in Subsection~\ref{sec:GHPF} to extract the high-frequency components of the graph signal, and then threshold it. If one or more of the Fourier transform coefficients defined in (\ref{equ:filteringGFT}) exceeds the threshold value, we conclude that there is an attack. That is, the maximum absolute element of $\mathcal{F}_\text{R}\{\widehat{\psivec}^\text{R}\}$,
 \begin{equation}
\label{equ:maxPsi}
\Psi_\text{R}^\text{R} =\left| \left| \mathcal{F}_\text{R}\{\widehat{\psivec}^\text{R}\}\right| \right|_{\infty},
\end{equation} 
is compared to a threshold, $\tau$. Two alternative procedures to set the threshold are described in the following: 
\subsubsection{Maximum Threshold}
In \cite{sandryhaila2014DSPonGraph} the threshold is defined as the maximum from a set of historic states:
 \begin{equation}
\label{equ:tauMax}
\tau_{\max} \define \max \{ \Psi_1, \Psi_2, \dots,  \Psi_{N}\},
\end{equation}
where $N$ is the number of historic states. Transferred to the application in power systems, the detection threshold $\tau_{\max}$ is thus the maximum of all $\Psi_\text{R}^\text{R}$ that can be calculated according to Algorithm~\ref{alg:FDIdetectionGeneral} based on a list of historic grid states $\left[ \widehat{\vvec}_1^\text{R}, \widehat{\vvec}_2^\text{R}, \dots,  \widehat{\vvec}_N^\text{R}\right]$. 
\subsubsection{Averaged threshold}
In \cite{silva2016PlantWideFaultDetection} the threshold is defined as the average of historic states plus a deviation term to set the size of the confidence interval as follows:
 \begin{equation}
\label{equ:tauStat}
\tau_\text{R}^\text{R} = \frac{1}{N} \sum_{k=1}^{N} \left(\Psi_\text{R}^\text{R}\right)_k + \sigma_{\Psi}^\text{R} \alpha_{\sigma},
\end{equation}
where $\sigma_{\Psi}^\text{R}$ is the standard deviation between the $\left(\Psi_\text{R}^\text{R}\right)_k$. The parameter $\alpha_{\sigma}$ is a tuning parameter that allows the choice of the confidence interval. 

The averaged threshold from (\ref{equ:tauStat}) is statistically more robust to outliers than the maximum threshold from (\ref{equ:tauMax}). Consistently, our simulations  showed that, indeed, the threshold value from (\ref{equ:tauStat}) led to better results than the one from (\ref{equ:tauMax}). Therefore, in the case studies in Section~\ref{sec:CaseStudy}, the definition of $\tau_\text{R}^\text{R}$ in (\ref{equ:tauStat}) is applied. 

In order to use the total variation, $S \left( \widehat{\vvec}^\text{R} \right)$, as an indicator for the presence of FDI attacks, the detection threshold is constructed by sustituting $\Psi_\text{R}^\text{R}$ by the smoothness of the historical states, $S \left( \widehat{\vvec}^\text{R} \right)$ in (\ref{equ:tauStat}):
 \begin{equation}
\label{equ:tauStatVariation}
\tau_{S}^\text{R} = \frac{1}{N} \sum_{k=1}^{N} \left[ S \left( \widehat{\vvec}^\text{R}_k \right)\right] + \sigma_{S}^\text{R} \alpha_{\sigma,S},
\end{equation}
using the mean value of all total variations derived from the historic grid states according to (\ref{equ:smoothnes}), and where $\sigma_{S}^\text{R}$ is the standard deviation between the variations of different historical sates scaled by the parameter $\alpha_{\sigma,S}$ that defines the confidence interval. 

The precision of the propsed FDI attack detection method strongly depends on the choice of the cutoff frequency, $\lambda_\text{cut}$. If the frequency band extends too far into the low-frequency part of the spectrum, normal grid states can also contribute with a high amplitude. On the other hand, if the frequency band is too small, attacks might pass undetected. In this work, the cutoff frequency is chosen such that for undisturbed states the contribution that is given by Fourier components above this frequency is smaller than \unit[0.1]{\%}. To find the threshold, we rely on concepts of graph spectral compression \cite{zhu2012GraphSpectralCompressedSensing}. There, the approximation error $\epsilon_{\gamma}^\text{R}$ that results from cutting off all high-frequency components above the $\gamma$-th frequency is defined as
 \begin{equation}
\label{equ:cutOffError}
\epsilon_{\gamma}^\text{R} = \sum_{m=\gamma}^{M} \left|\mathcal{F}_\text{R}\{\widehat{\vvec}^\text{R}\}\left( \lambda_m \right) \right|^2.  
\end{equation}
In \cite{zhu2012GraphSpectralCompressedSensing}, this equation is used to compress and approximate the signal. In this work, we fix $\epsilon_{\gamma}^\text{R}$ to a desired value and iteratively decrease $\gamma$ until further decrease would lead to the situation where the right hand side of (\ref{equ:cutOffError}) would be larger than the fixed $\epsilon_{\gamma}^\text{R}$. The cutoff frequency then is given by  $\lambda_\gamma$. 

Algorithm~\ref{alg:FDIdetectionGeneral} summarizes the proposed approach of applying a GHPF on an arbitrary signal to obtain the maximum norm of its high-frequency components. The overall proposed detection procedure is summarized in Algorithm~\ref{alg:FDIs}. This algorithm consists of applying Algorithm~\ref{alg:FDIdetectionGeneral} on all four terms of (\ref{equ:ACdecompose}).

{\SetAlgoNoLine
\begin{algorithm}[t]
   \SetKwInOut{Input}{Input} 
   	 \Input{ 
\begin{minipage}[t]{2.6in}
\begin{itemize}
       \item real Laplacian matrix $\Lmat$
       \item state vector $\svec$
       \item maximal approximation error $\epsilon_{\gamma}$
       \item list of historic states $\left[\svec_1, \svec_2, \dots, \svec_N\right]$
       \item confidence interval $\alpha_{\sigma}$
   \end{itemize}
   \end{minipage}	
   }
   \vspace{2mm}
   \SetKwInOut{Output}{Output}
   \Output{$\Psi$, $\tau$}
    \begin{enumerate}
    \item calculate eigenvalue decomposition of $\Lmat = \Umat \Lambdamat \Umat^T$ to get the matrix of eigenvectors $\Umat$ and the eigenvalues $\lambda_{1}, \lambda_{2}, \dots, \lambda_{M}$
    \item calculate GFT $\mathcal{F}\{\svec \} = \Umat^T \svec$
    \item construct polynomial matrix $\Mmat$ where $\Mmat_i = \left[1, \lambda_i,  \dots ,\lambda_i^L\right]$
    \item find cutoff frequency $\lambda_\text{cut}$ based on maximal approximation error $\epsilon_{\gamma} = \sum_{m=\gamma}^{M} \left|\mathcal{F}\{\svec \} \left( \lambda_m \right) \right|^2$
    \item solve linear equation $ \Mmat \hvec = \onevec_{\lambda > \lambda_\text{cut}}$ to find $\hvec = \left[ h_0, h_1, \dots , h_L\right]^T$
    
    \item apply filter $\mathcal{F}\{ \tilde{\svec}\}= h \left(\Lambdamat\right)\mathcal{F}\{\svec\}$
    \item calculate maximum norm $\Psi =\left| \left| \mathcal{F}\{ \tilde{\svec}\} \right| \right|_{\infty}$
    \item based on historic states, calculate detection threshold $\tau = \mu_{\Psi} + \sigma_{\Psi} \alpha_{\sigma}$, where $\Psi_k =\left| \left| 
    h \left(\Lambdamat\right)\mathcal{F}\{\svec_k \} \right| \right|_{\infty} $ and $\mu_{\Psi}$ is the mean and $\sigma_{\Psi}$ the standard deviation of all $\Psi_k$
    \item return $\Psi$, $\tau$
   \end{enumerate}
   \caption{GFT and GHPF design for general input.}
   \label{alg:FDIdetectionGeneral}
\end{algorithm}
}

{\SetAlgoNoLine
\begin{algorithm}[t]
   \SetKwInOut{Input}{Input} 
   	 \Input{ 
\begin{minipage}[t]{3in}
\begin{itemize}
       \item admittance matrix of the grid in the form of a Laplacian matrix $\Ymat$
       \item vector of estimated bus voltages $\widehat{\vvec}$
       \item maximal error $\epsilon_{\gamma}^\text{R}$, $\epsilon_{\gamma}^\text{J}$
       \item list of historic grid states $\left[ \widehat{\vvec}_1, \widehat{\vvec}_2, \dots, \widehat{\vvec}_N\right]$
       \item confidence interval $\alpha_{\sigma}$
   \end{itemize}
   \end{minipage}	
   }
   \vspace{2mm}
   \SetKwInOut{Output}{Output}
   \Output{$\mathcal{H}_0$ or $\mathcal{H}_1$}
    \begin{enumerate}
    \item $\Ymat^\text{R} = \text{Re} \left( \Ymat \right)$  and $\Ymat^\text{J} = - \text{Im} \left( \Ymat \right)$
    \item $\widehat{\vvec}^\text{R} = \text{Re} \left( \widehat{\vvec} \right)$  and $\widehat{\vvec}^\text{J} = \text{Im} \left( \widehat{\vvec} \right)$ 
    
    \item apply Algorithm~\ref{alg:FDIdetectionGeneral} on $\Ymat^\text{R}, \widehat{\vvec}^\text{R}, \epsilon_{\gamma}^\text{R}, \left[ \widehat{\vvec}^\text{R}_1,
    \widehat{\vvec}^\text{R}_2, \dots,  \widehat{\vvec}^\text{R}_N\right],$ $\alpha_{\sigma}$ to obtain $\Psi_\text{R}^\text{R}$, $\tau_\text{R}^\text{R}$
    
     \item apply Algorithm~\ref{alg:FDIdetectionGeneral} on $\Ymat^\text{J}, \widehat{\vvec}^\text{R}, \epsilon_{\gamma}^\text{R}, \left[ \widehat{\vvec}^\text{R}_1,
    \widehat{\vvec}^\text{R}_2, \dots,  \widehat{\vvec}^\text{R}_N\right],$ $\alpha_{\sigma}$ to obtain $\Psi_\text{J}^\text{R}$, $\tau_\text{J}^\text{R}$
    
    \item apply Algorithm~\ref{alg:FDIdetectionGeneral} on $\Ymat^\text{J}, \widehat{\vvec}^\text{J}, \epsilon_{\gamma}^\text{J}, \left[ \widehat{\vvec}^\text{J}_1, \widehat{\vvec}^\text{J}_2, \dots,  \widehat{\vvec}^\text{J}_N\right],$ \\$\alpha_{\sigma}$ to obtain $\Psi_\text{J}^\text{J}$, $\tau_\text{J}^\text{J}$
    
    \item apply Algorithm~\ref{alg:FDIdetectionGeneral} on $\Ymat^\text{R}, \widehat{\vvec}^\text{J}, \epsilon_{\gamma}^\text{J}, \left[ \widehat{\vvec}^\text{J}_1,
    \widehat{\vvec}^\text{J}_2, \dots,  \widehat{\vvec}^\text{J}_N\right],$ \\ $\alpha_{\sigma}$ to obtain $\Psi_\text{R}^\text{J}$, $\tau_\text{R}^\text{J}$
    
    \item \eIf{$(\Psi_\text{R}^\text{R} >\tau_\text{R}^\text{R}) \lor (\Psi_\text{J}^\text{J}>\tau_\text{J}^\text{J}) \lor (\Psi_\text{R}^\text{J} >\tau_\text{R}^\text{J}) \lor (\Psi_\text{J}^\text{R}>\tau_\text{J}^\text{R})$}
	  {
	\Indp      return $\mathcal{H}_1$
	  }
	  {
	\Indp      return $\mathcal{H}_0$
	  }
   \end{enumerate}
   \caption{FDI attack detection for AC model based on GSP.}
   \label{alg:FDIs}
\end{algorithm}
}

\section{Case Study}
\label{sec:CaseStudy}
In this section, the proposed approach is applied and investigated on the IEEE 14-bus test grid. This test grid is often used in the context of FDI attack detection \cite{kosut2011MaliciousDataAttacks, jiang2017DefenseMechanisms, ashok2018OnlineDetectionStealthy} and attack construction \cite{yu2015FalseDataPCA, liu2017FalseDataAttackAC}. The voltage angles and magnitudes of this grid are calculated from the AC power flow solver implemented by the pandapower-tool \cite{thurner2018Pandapower}. Bus number 1 is set to be the slack bus. In the following, several different test cases are investigated. In Subsection~\ref{sec:TestCase1} the total variation of undisturbed grid states is investigated. Subsection~\ref{sec:TestCase2} investigates attacks on the voltage and magnitude separately, while applying different threshold definitions. In Subsection~\ref{sec:TestCaseError} we investigate the robustness of our method against state estimation noise, and in Subsection~\ref{sec:TestCase4} a constructed undetectable FDI attack that combines attacks on the angle and magnitude is launched and detected. Finally, in Subsection~\ref{sec:Comparison} the performance of the proposed method is compared with that of previous works.

\subsection{Test Case 1: Total Variation of Undisturbed Grid States}
\label{sec:TestCase1}
In this test case, the total variation of the standard load situation is calculated using (\ref{equ:smoothnes}), normalized by the number of buses in the grid, $M$, and shown in Table~\ref{tab:IEEEgridSmoothness} for IEEE 14-bus, 24-bus, 30-bus, and 118-bus test grids \cite{thurner2018Pandapower}. It can be seen that the real part of the graph signal has a smaller total variation, and thus is smoother than the imaginary part.

\begin{table}[!t]
\renewcommand{\arraystretch}{1.3}
\caption{Smoothness measured as total variation of standard load situation in several IEEE test cases.}
\label{tab:IEEEgridSmoothness}
\centering
\begin{tabular}{l|c|c}
& $ \frac{S \left( \widehat{\vvec}^\text{R} \right)}{M}$ & $\frac{S \left( \widehat{\vvec}^\text{J} \right)}{M}$  \\
\hline
IEEE-14 & 0.13 & 3.48 \\
\hline
IEEE-24 & 0.16 & 17.98\\
\hline
IEEE-30 & 0.04 & 0.18\\
\hline 
IEEE-118 & 0.11 & 5.82 \\
\end{tabular}
\end{table}

\begin{figure}[!t]
\centering
\includegraphics[width=2.9in]{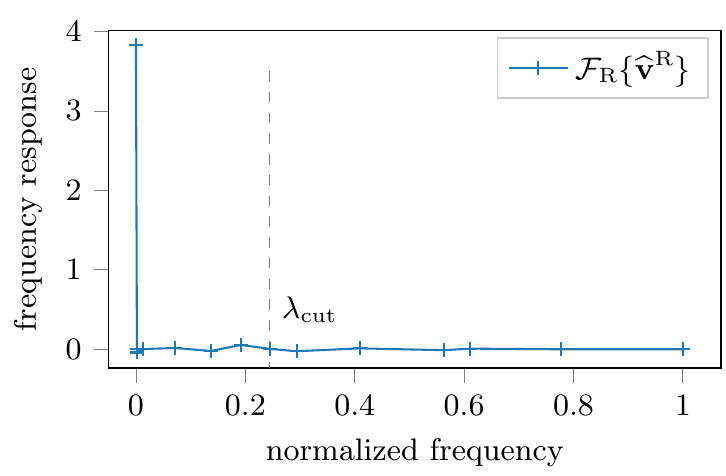}
\caption{Frequency response of the real part, $\mathcal{F}_\text{R}\{\widehat{\vvec}^\text{R}\}$, of a valid grid state.}
\label{fig:fResponse_R}
\end{figure}

\begin{figure}[!t]
\centering
\includegraphics[width=2.9in]{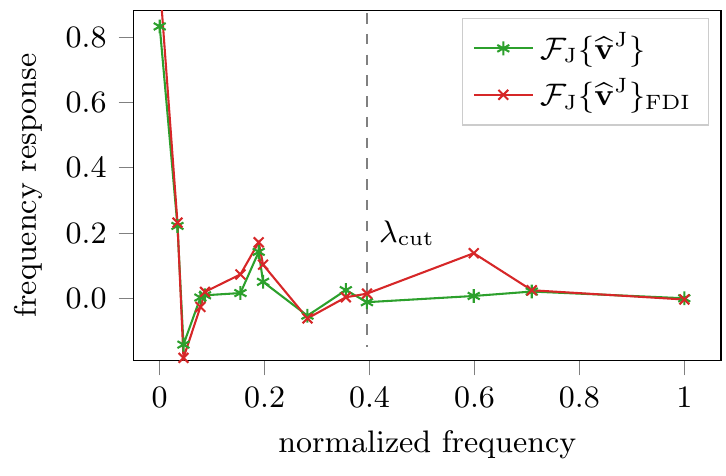}
\caption{Frequency response of the imaginary part, $\mathcal{F}_\text{J}\{\widehat{\vvec}^\text{J}\}$, of a valid grid state, as well as the frequency response of a state subject to an FDI attack, $\mathcal{F}_\text{J}\{\widehat{\vvec}^\text{J}\}_\text{FDI}$.}
\label{fig:fResponse_I}
\end{figure}

For the IEEE 14-bus test grid, the smoothness of the real and imaginary parts, $S \left( \widehat{\vvec}^\text{R} \right)$ and $S \left( \widehat{\vvec}^\text{J} \right)$, are visualized by plotting the frequency response calculated using (\ref{equ:GFT}) of the two signals $\mathcal{F}_\text{R}\{\widehat{\vvec}^\text{R}\}$ and $\mathcal{F}_\text{J}\{\widehat{\vvec}^\text{J}\}$, as shown in Fig.~\ref{fig:fResponse_R} and Fig.~\ref{fig:fResponse_I}. These figures validate the decaying characteristic of the Fourier components, and show that $\mathcal{F}_\text{R}\{\widehat{\vvec}^\text{R}\}$ is smoother than $\mathcal{F}_\text{J}\{\widehat{\vvec}^\text{J}\}$. Only the contributions lying above the cutoff frequency $\lambda_\text{cut}$ pass the high-pass filter and are used for the detection. In Subsection~\ref{sec:GFT} it is claimed that FDI attacks destroy this decaying behavior. This is visualized, for example, in Fig.~\ref{fig:fResponse_I} with $\mathcal{F}_\text{J}\{\widehat{\vvec}^\text{J}\}_\text{FDI}$ for an angle attack of $\unit[10]{^\circ}$ on bus number~9. In this figure, and throughout, the graph frequencies are normalized: $\lambda_i = \frac{\lambda_i}{\left|\lambda_\text{max}\right|}, i= 1, 2,\dots, M$.

\subsection{Test Case 2: Voltage Angle and Magnitude Attack}
\label{sec:TestCase2}
In this test case, the effects of undetectable sparse attacks are systematically investigated for attacks on the voltage magnitude and angle separately on all buses of the grid (except the slack bus), $k = 2, 3, \dots, M$. The construction of such an attack on the angle is described in \cite{kosut2011MaliciousDataAttacks}, and, similarly, the description given in \cite{hug2012VulnerabilityAssessmentAC} is used to construct an attack on the voltage magnitude. In particular, the attack is realized by modifying either the magnitude $V_k$ or the angle $\varphi_k$ according to (\ref{equ:xbad}) by adding a certain offset $\widehat{v}_{k,\text{FDI}} = \widehat{v}_k + c$ with $c =  A_k e^{j a_k}$, where the attack on the angle is $a_k$ and the attack on the magnitude is $A_k$. Specifically, $a_k \in \left[ -12, -11, \dots, 12 \right]$ degrees and $A_k \in \left[ -0.2, -0.18,\dots, 0.2 \right]$ p.u.. Every attack is executed by 100 Monte Carlo simulations of randomly generated grid states following
\begin{equation}
\label{equ:randomLoad}
 \begin{aligned}
P_k & = P^0_k \left|y_\text{P} \right| \\
Q_k & = Q^0_k \left|y_\text{Q} \right| \\
\end{aligned},
\end{equation}
where $y_\text{P}, y_\text{Q}$ are Gaussian random variables with mean 1 and variance $\sigma^2$. The terms $P^0_k$ and $Q^0_k$ are the nominal active and reactive loads, respectively, for bus $k$, as given in the test case. For the GFT and subsequent filtering, the cutoff frequency, $\lambda_\text{cut}$, is found by allowing an approximation error of $\epsilon_{\gamma}^\text{R} = 0.0001$ and $\epsilon_{\gamma}^\text{J} = 0.001$ based on (\ref{equ:cutOffError}) and the approach explained there. As the graph is smoother for the real than for the imaginary part, as shown in Table~\ref{tab:IEEEgridSmoothness}, the error is chosen to be smaller for the real part than for the imaginary part. 

To generate the required historic data, the loads of the grid are again randomly scaled following (\ref{equ:randomLoad}). Then, an AC power that provides the grid states, $\left[ \widehat{\vvec}_1, \widehat{\vvec}_2, \dots,  \widehat{\vvec}_N\right]$ is performed. 

\begin{figure}[!t]
\centering
\includegraphics[width=3.1in]{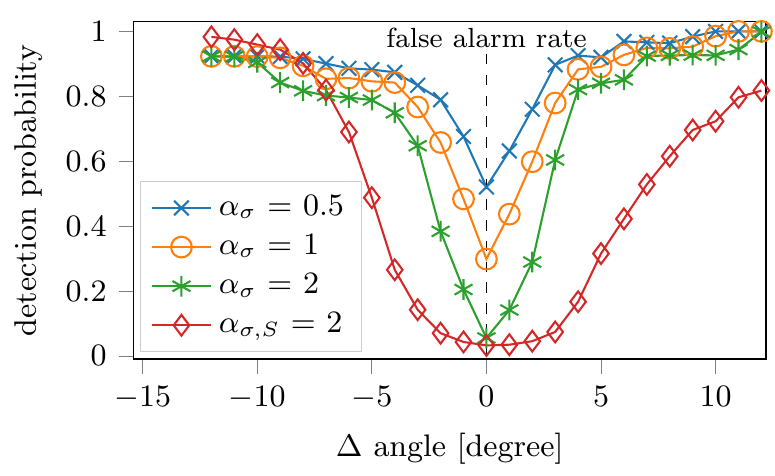}
\caption{Test case 2: Detection probability for FDI attacks that exceed a particular change of angle. The size of the confidence interval is scaled with $\alpha_\sigma$, as defined in (\ref{equ:tauStat}).}
\label{fig:detectionProbabilityAngle}
\end{figure}

\begin{figure}[!t]
\centering
\includegraphics[width=3.1in]{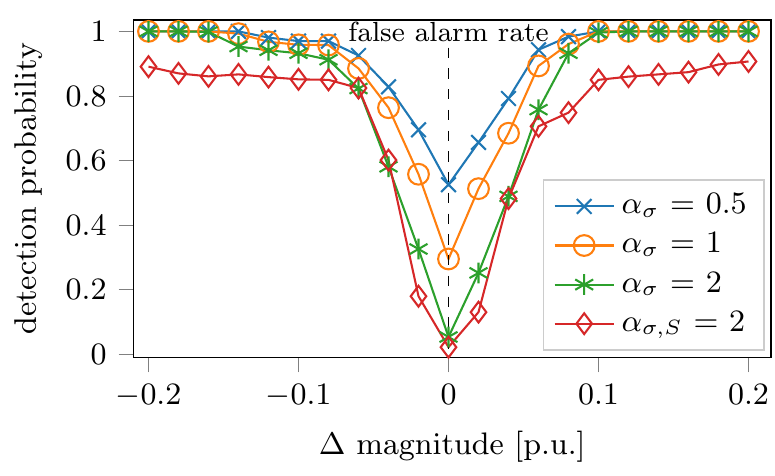}
\caption{Test case 2: Detection probability for FDI attacks that exceed a particular change of magnitude. The size of the confidence interval is scaled with $\alpha_\sigma$, as defined in (\ref{equ:tauStat}).}
\label{fig:detectionProbabilityMag}
\end{figure}

The results of these tests are summarized in Figs.~\ref{fig:detectionProbabilityAngle} and \ref{fig:detectionProbabilityMag} for attacks on the angle and on the magnitude, respectively. The curves related to $\alpha_{\sigma,S}$ show the detection based on the smoothness threshold from (\ref{equ:tauStatVariation}), while the others show the detection based on GFT from (\ref{equ:tauStat}) for various values of $\alpha_{\sigma}$. These figures show the probability that an FDI attack, with an impact exceeding a particular angle $a_k$ or magnitude $A_k$, is detected by the proposed method for any attacked bus. 
It should be noted that the detection probability at the point of $\Delta = 0$, i.e. when there is no attack, gives the false alarm rate. Thus, the choice of $\alpha_{\sigma}$ from (\ref{equ:tauStat}) directly determines the false alarm rate. For example, in Fig.~\ref{fig:detectionProbabilityAngle} the false alarm probabilities are roughly 0.5, 0.3 and  0.05 for $\alpha_\sigma = 0.5, 1, 2$. According to (\ref{equ:tauStat}), $\alpha_{\sigma}$ scales the confidence interval around the mean value. Thus, the false alarm rate is of the order of magnitude of the percentage of values of a normal distribution that lie outside the confidence interval.  

In Table~\ref{tab:detectionPercent} the contributions of the different terms of (\ref{equ:ACdecompose}) to the attack detection are listed. It can be seen from this that both the analysis related to the real part of the state vector, $\Psi_\text{R}^\text{R}$ and $\Psi_\text{J}^\text{R}$, as well as to the imaginary part, $\Psi_\text{J}^\text{J}$ and $\Psi_\text{R}^\text{J}$, contribute to the detection. Attacks on the angle are mostly detected by the imaginary part of the voltage vector, while attacks on the voltage magnitude are, largely, detected by the real part of the voltage vector. This behavior is explained in Subsection~\ref{sec:ContributionRealImaginary}. 
\begin{table}[!t]
\renewcommand{\arraystretch}{1.3}
\caption{Percentage of detected attacks that are detected by the real or the imaginary part of the state vector.}
\label{tab:detectionPercent}
\centering
\begin{tabular}{l|c|c|c}
& $\left(\Psi_\text{R}^\text{R} \lor \Psi_\text{J}^\text{R}\right) \land \left(\Psi_\text{J}^\text{J} \lor \Psi_\text{R}^\text{J}\right)$ & $\Psi_\text{R}^\text{R} \lor \Psi_\text{J}^\text{R}$ & $\Psi_\text{J}^\text{J} \lor \Psi_\text{R}^\text{J}$  \\
\hline
Angle attack & $\unit[17]{\%}$ & $\unit[18]{\%}$ & $\unit[65]{\%}$ \\
\hline
Magnitude attack & $\unit[1]{\%}$ & $\unit[97]{\%}$ & $\unit[2]{\%}$\\
\end{tabular}
\end{table}


\subsection{Test Case 3: Robustness against State Estimation Errors}
\label{sec:TestCaseError}
\begin{figure}[!t]
\centering
\includegraphics[width=3.1in]{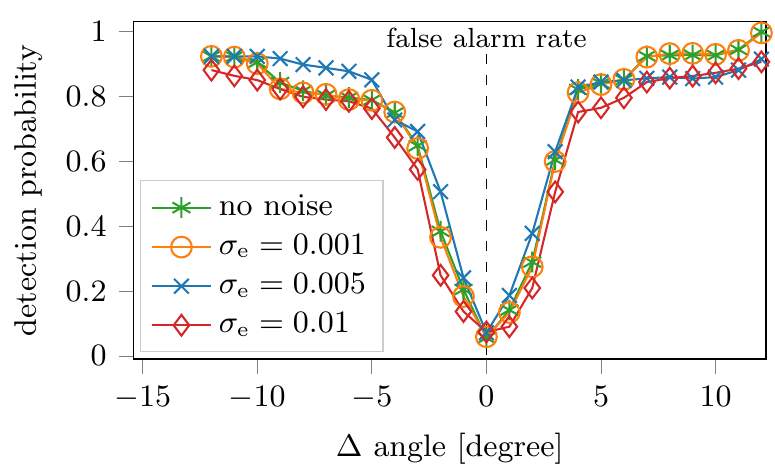}
\caption{Test case 3: Detection probability for FDI attacks that exceed a particular change of angle with varying noise variance.}
\label{fig:detectionProbabilityAngleNoise}
\end{figure}

\begin{figure}[!t]
\centering
\includegraphics[width=3.1in]{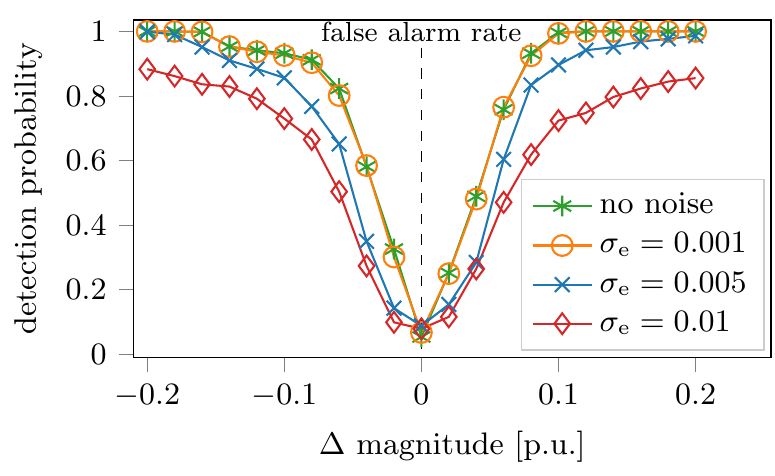}
\caption{Test case 3: Detection probability for FDI attacks that exceed a particular change of magnitude with varying noise variance.}
\label{fig:detectionProbabilityMagNoise}
\end{figure}

According to (\ref{equ:xbad}), the estimated grid state, $\widehat{\vvec}$, is influenced not only by a possible attack vector, $\cvec$ but also by random noise, $\evec$, that results from the measurement errors in the state estimation. In order to validate the robustness of the proposed method against this noise term, we added noise vectors with varying standard deviations, $\sigma_\text{e}$, to the voltage vector before simulating the attacks as described in the previous test case. Since the standard deviations of noise errors in PSSE are often considered to be around $\sigma_\text{e}=0.001$ \cite{ashok2018OnlineDetectionStealthy}, in our simulations we assumed standard deviations of $\sigma_\text{e}=0.001, 0.005, 0.01$. The results of these simulations can be found in Figs.~\ref{fig:detectionProbabilityAngleNoise} and \ref{fig:detectionProbabilityMagNoise}.
These figures show that the proposed detection method is, in fact very, robust against state estimation noises, even for standard deviations that are larger as generally assumed.

\subsection{Test Case 4: Wide Scale Undetectable FDI Attack}
\label{sec:TestCase4}
\begin{figure}[!t]
\centering
\includegraphics[width=3.1in]{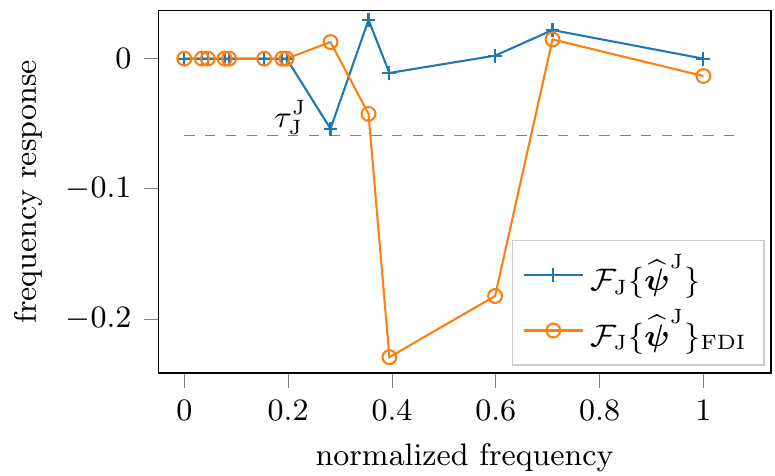}
\caption{Test case 4: Frequency response of $\mathcal{F}_\text{J}\{\widehat{\psivec}^\text{J}\}$ with and without undetectable FDI attack. For the attacked case, $\mathcal{F}_\text{J}\{\widehat{\psivec}^\text{J}\}_\text{FDI}$, the threshold $\tau_\text{J}^\text{J}$ is exceeded.}
\label{fig:fResponseComplexI}
\end{figure}

This test case relies on the constructed undetectable FDI attack proposed in \cite{liu2017FalseDataAttackAC}. Based on the IEEE-14 bus sytem, the authors provide two data sets, one for an undisturbed case and one for an attack on the buses 6 and 9 - 14. Both the angle and the magnitude of these bus voltages have been tampered with. For the detection, the same parameters are used as in the previous test cases with $\alpha_{\sigma} = 2$. Fig.~\ref{fig:fResponseComplexI} shows the Fourier transformed and filtered components of the imaginary part of the voltage $\mathcal{F}_\text{J}\{\widehat{\psivec}^\text{J}\}$,  both for the undisturbed case and for the undetectable FDI attack, based on the imaginary part of the Laplacian matrix, $\Ymat^\text{J}$. For the latter, the detection threshold $\tau_\text{J}^\text{J}$ is exceeded and, thus, the undetectable FDI attack is detected. 

\subsection{Comparison with other Approaches}
\label{sec:Comparison}
The aim of this subsection is to compare the behavior of previous works with the results we obtained in our case study in Subsection~\ref{sec:TestCase2}. 

 The detection method in \cite{kosut2011MaliciousDataAttacks} is based on an ``energy residue'' heuristic. In order to compare this method with the method proposed herein, we implemented it comparing the norm of the grid state, $\left \lVert \widehat{\vvec}_t \right \rVert$ , with a threshold, which was set such that the false alarm rate is equal to the outcome of the proposed method.

\begin{figure}[!t]
\centering
\includegraphics[width=3.2in]{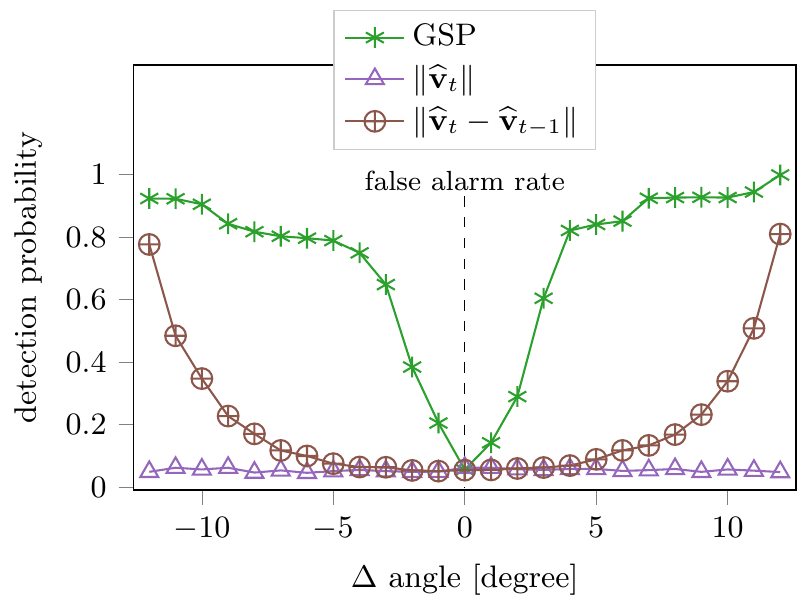}
\caption{Comparison between the proposed method based on GSP and two previously published methods, one based on the norm of the grid state,$\left \lVert \widehat{\vvec}_t \right \rVert$ and the other based on the norm of the residual between two consecutive grid states, $\left \lVert \widehat{\vvec}_t - \widehat{\vvec}_{t-1} \right \rVert$.}
\label{fig:compMethod}
\end{figure}

We also implemented an adapted version of the work proposed in \cite{jiang2017DefenseMechanisms}, which is based on a residual evaluation between the grid states of two consecutive time steps, $\left \lVert \widehat{\vvec}_t - \widehat{\vvec}_{t-1} \right \rVert$. This residual is then compared with a threshold. We again chose the threshold such that the false alarm rate is equal to the outcome of the proposed method. Fig.~\ref{fig:compMethod} shows the comparison between the three detection methods for attacks on the voltage angle. It can be seen that the first method is unable to detect the attacks, while the second approach based on the residual between two time steps, only succeeds in detecting very strong attacks.

\section{Remarks}
\label{sec:Remarks}
In this section, we discuss some aspects and characteristics of the proposed method that emerged from the case studies. In Subsection~\ref{sec:DCmodel} we also summarize the application of the FDI attack detection method on the DC model. 

\subsection{Detection Characteristics}
\label{sec:detectionLimit}
The case studies in the previous section demonstrate that we are able to detect previously undetectable attacks. However, it can be verified that attacks on some of the buses are better detected than attacks on others, even if they have the same impact. This behavior can be explained by looking at the diagonal elements of $\Ymat$. As given in (\ref{equ:YForm}), the diagonal elements in the Laplacian matrix, $\Ymat_{k,k}$, sum up the weights of all edges connected to node $k$. The smaller this value is, the more ``loosely'' coupled is the bus. For such buses, even large differences between
$\widehat{v}_k$ and $\widehat{v}_l$ do not contribute significantly to the total variation as it is defined in (\ref{equ:smoothnes}). In summary, for buses with small diagonal values in $\Ymat$ compared to the other buses, attacks are the most difficult to detect. 

For detection based on the smoothness, the boundary condition for detectability is given by (\ref{equ:smoothnes}). As long as the attack does not increase $S$, it cannot be detected. 

\subsection{Contribution of Real and Imaginary Parts}
\label{sec:ContributionRealImaginary}
Table~\ref{tab:detectionPercent} summarizes the second case study and shows that attacks on the magnitude are mainly detected by $\Psi_\text{R}^\text{R}$ or $\Psi_\text{J}^\text{R}$, both of which are related to the real part of the voltage vector. Attacks on the angle are mainly detected by $\Psi_\text{J}^\text{J}$ or $\Psi_\text{R}^\text{J}$ that are related to the imaginary part of the voltage vector. This behavior can be explained by writing the complex voltage in the Cartesian and the Euler formulations:
\begin{equation}
\label{equ:vcomplex}
\vvec = \vvec^\text{R} + j \vvec^\text{J} =  \Vmat \cos \varPhivec +  j \Vmat \sin \varPhivec.
\end{equation}
Applying the small angle approximation in (\ref{equ:vcomplex}), i.e. $\sin \varphi \approx \varphi$ and $\cos \varphi \approx 1$, leads to the following relation:
\begin{equation}
\label{equ:smallAngle}
\vvec \approx \Vmat + j \varPhivec. 
\end{equation}
Thus, FDI attacks on the magnitude mainly influence the real part and thus, can be best detected by the real part, while FDI attacks on the angle mainly affect the imaginary part of the voltage and thus, can be best detected by the imaginary part.

\subsection{FDI Attack Detection for the DC Model}
\label{sec:DCmodel}
The DC power flow model is based on the following assumptions \cite{wood1984PowerGenerationOperation}:
\begin{itemize}
 \item voltage magnitude is fixed at 1 p.u.,
 \item each line $ e_{k,l} \in E$ is characterized by its reactance, neglecting its resistance, 
 \begin{equation}
 \label{equ:admittanceDC}
y_{k,l}=\frac{1}{x_{k,l}},~\forall e_{k,l} \in E,
\end{equation}
\item the angle difference between two connected buses is quite small.
\end{itemize}
Under these assumptions, the vector of bus voltage angles, $\varPhivec = \left[\varphi_1, \varphi_2, \dots, \varphi_{M} \right]^T \in \R^M$, is linked to the vector of real power injection or consumption of the buses, ${\pvec = \left[P_1, P_2, \dots, P_{M}\right]^T \in \R^M}$, according to 
\begin{equation}
\label{equ:PowerFlowDC}
\pvec = \Ymat_\text{DC} \varPhivec,
\end{equation}
where $\Ymat_\text{DC} \in \R^{M \times M}$ is the admittance matrix for the DC model. In this case, the matrix $\Ymat_\text{DC}$ is a real, symmetric, positive-semidefinite matrix, which takes the form of a weighted Laplacian matrix, as defined in (\ref{equ:YForm}), with the elements defined in (\ref{equ:admittanceDC}). The proposed approach for FDI attack detection can be applied to the DC model. The details and a full algorithm for the DC model case can be found in our preliminary work \cite{drayer2018BadDataGSP}. 

\section{Conclusion}
\label{sec:Conclusion}
In this paper we present a novel method for the detection of FDI attacks in power systems, which relies on the inherent graph structure of the grid. The proposed method uses the AC model as a basis to describe the electrical behavior of the grid and its associated graph representation in the form of a Laplacian matrix. The method is based on performing a GFT and filtering the high-frequency components associated with the large eigenvalues of the Laplacian matrix. Large contributions in the high-frequency range indicate the existence of anomalies or malicious FDI attacks. Extensive case studies show that the graph signals originating from power systems exhibit the required decaying behavior in their Fourier components. This concentration within the low-frequency components is destroyed for grid states originating from FDI attacks. This facilitates the detection of previously undetectable attacks based on the high-frequency content.

\ifCLASSOPTIONcaptionsoff
  \newpage
\fi


\bibliographystyle{IEEEtran}
\bibliography{IEEEabrv,literatur_paper}

\begin{thebibliography}{10}
\providecommand{\url}[1]{#1}
\csname url@samestyle\endcsname
\providecommand{\newblock}{\relax}
\providecommand{\bibinfo}[2]{#2}
\providecommand{\BIBentrySTDinterwordspacing}{\spaceskip=0pt\relax}
\providecommand{\BIBentryALTinterwordstretchfactor}{4}
\providecommand{\BIBentryALTinterwordspacing}{\spaceskip=\fontdimen2\font plus
\BIBentryALTinterwordstretchfactor\fontdimen3\font minus
  \fontdimen4\font\relax}
\providecommand{\BIBforeignlanguage}[2]{{%
\expandafter\ifx\csname l@#1\endcsname\relax
\typeout{** WARNING: IEEEtran.bst: No hyphenation pattern has been}%
\typeout{** loaded for the language `#1'. Using the pattern for}%
\typeout{** the default language instead.}%
\else
\language=\csname l@#1\endcsname
\fi
#2}}
\providecommand{\BIBdecl}{\relax}
\BIBdecl

\bibitem{nato2016CriticalInfrastructureProtection}
A.~Niglia, \emph{Critical Infrastructure Protection Against Hybrid Warfare
  Security Related Challenges.}, ser. NATO Science for Peace and Security
  Series, D, Information and Communication Security.\hskip 1em plus 0.5em minus
  0.4em\relax IOS Press, 2016, vol.~46.

\bibitem{lewis2006CIProtection}
T.~G. Lewis, \emph{\BIBforeignlanguage{eng}{Critical infrastructure protection
  in homeland security: defending a networked nation}}.\hskip 1em plus 0.5em
  minus 0.4em\relax Hoboken, N.J.: Wiley, 2006.

\bibitem{buldyrev2010CatastrophicCascade}
S.~V. Buldyrev, R.~Parshani, G.~Paul, H.~E. Stanley, and S.~Havlin,
  ``Catastrophic cascade of failures in interdependent networks,''
  \emph{Nature}, vol. 464, no. 7291, p. 1025–1028, 2010.

\bibitem{teixeira2015SecureControlSystems}
A.~Teixeira, K.~C. Sou, H.~Sandberg, and K.~H. Johansson, ``Secure control
  systems: A quantitative risk management approach,'' \emph{IEEE Control
  Systems}, vol.~35, no.~1, p. 24–45, Feb. 2015.

\bibitem{smith2016GoingBeyondCybersecurity}
E.~Smith, S.~Corzine, D.~Racey, P.~Dunne, C.~Hassett, and J.~Weiss, ``Going
  beyond cybersecurity compliance: What power and utility companies really need
  to consider,'' \emph{IEEE Power Energy Mag.}, vol.~14, no.~5, p. 48–56,
  Sep. 2016.

\bibitem{soltan2018CyberPhysicalAttack}
S.~Soltan, M.~Yannakakis, and G.~Zussman, ``Power grid state estimation
  following a joint cyber and physical attack,'' \emph{IEEE Trans. Control of
  Network Syst.}, vol.~5, no.~1, p. 499–512, Mar. 2018.

\bibitem{drayer2018DistributedSelfHealing}
E.~Drayer, N.~Kechagia, J.~Hegemann, M.~Braun, M.~Gabel, and R.~Caire,
  ``Distributed self-healing for distribution grids with evolving search
  space,'' \emph{IEEE Trans. Power Del.}, vol.~33, no.~4, p. 1755–1764, Aug.
  2018.

\bibitem{routtenberg2013JointFrequency}
T.~Routtenberg and L.~Tong, ``Joint frequency and phasor estimation under the
  {KCL} constraint,'' \emph{IEEE Signal Process. Lett.}, vol.~20, no.~6, p.
  575–578, Jun. 2013.

\bibitem{routtenberg2017PUMBasedDetection}
T.~Routtenberg, R.~Concepcion, and L.~Tong, ``{PMU}-based detection of voltage
  imbalances with tolerance constraints,'' \emph{IEEE Trans. Power Del.},
  vol.~32, no.~1, p. 484–494, Feb. 2017.

\bibitem{routtenberg2018CentralizedIdentification}
T.~Routtenberg and Y.~C. Eldar, ``Centralized identification of imbalances in
  power networks with synchrophasor data,'' \emph{IEEE Trans. Power Syst.},
  vol.~33, no.~2, p. 1981–1992, Mar. 2018.

\bibitem{jin2018PowerGridACStateEstimation}
M.~Jin, J.~Lavaei, and K.~H. Johansson, ``Power grid {AC}-based state
  estimation: Vulnerability analysis against cyber attacks,'' \emph{IEEE Trans.
  Autom. Control}, 2018.

\bibitem{abur2004PowerSystemStateEstimation}
A.~Abur and A.~G. Expésito, \emph{Power system state estimation - Theory and
  implementation}.\hskip 1em plus 0.5em minus 0.4em\relax Marcel Dekker, 2004.

\bibitem{giannakis2013MonitoringOptimization}
G.~B. Giannakis, V.~Kekatos, N.~Gatsis, S.~Kim, H.~Zhu, and B.~F. Wollenberg,
  ``Monitoring and optimization for power grids: A signal processing
  perspective,'' \emph{IEEE Signal Process. Mag.}, vol.~30, no.~5, p.
  107–128, Sep. 2013.

\bibitem{liang2017ReviewFalseDataInjection}
G.~Liang, J.~Zhao, F.~Luo, S.~R. Weller, and Z.~Y. Dong, ``A review of false
  data injection attacks against modern power systems,'' \emph{IEEE Trans.
  Smart Grid}, vol.~8, no.~4, p. 1630–1638, Jul. 2017.

\bibitem{kim2015AgainstDataAttacks}
J.~Kim and L.~Tong, ``Against data attacks on smart grid operations: Attack
  mechanisms and security measures,'' in \emph{Cyber Physical Systems Approach
  to Smart Electric Power Grid}.\hskip 1em plus 0.5em minus 0.4em\relax
  Springer, 2015, p. 359–383.

\bibitem{liang2016VulnerabilityAnalysisFalseData}
J.~Liang, L.~Sankar, and O.~Kosut, ``Vulnerability analysis and consequences of
  false data injection attack on power system state estimation,'' \emph{IEEE
  Trans. Power Syst.}, vol.~31, no.~5, p. 3864–3872, Sep. 2016.

\bibitem{handschin1975BadData}
E.~Handschin, F.~C. Schweppe, J.~Kohlas, and A.~Fiechter, ``Bad data analysis
  for power system state estimation,'' \emph{IEEE Trans. Power App. Syst.},
  vol.~94, no.~2, p. 329–337, Mar. 1975.

\bibitem{monticelli1983ReliableBadData}
A.~Monticelli and A.~Garcia, ``Reliable bad data processing for real-time state
  estimation,'' \emph{IEEE Trans. Power App. Syst.}, vol. PAS-102, no.~5, p.
  1126–1139, May 1983.

\bibitem{liu2011FalseDataInjection}
Y.~Liu, P.~Ning, and M.~K. Reiter, ``False data injection attacks against state
  estimation in electric power grids,'' \emph{ACM Trans. Information and Syst.
  Security}, vol.~14, no.~1, 2011.

\bibitem{li2013BlindToplogyIdentification}
X.~Li, H.~V. Poor, and A.~Scaglione, ``Blind topology identification for power
  systems,'' in \emph{IEEE Int. Conf. Smart Grid Communications
  (SmartGridComm)}, Oct. 2013, p. 91–96.

\bibitem{anwar2016EstimationGridToplogy}
A.~Anwar, A.~Mahmood, and M.~Pickering, ``Estimation of smart grid topology
  using {SCADA} measurements,'' in \emph{IEEE Int. Conf. Smart Grid
  Communications (SmartGridComm)}, Nov. 2016, p. 539–544.

\bibitem{gera2017BlindEstimation}
I.~Gera, Y.~Yakoby, and T.~Routtenberg, ``Blind estimation of states and
  topology ({BEST}) in power systems,'' in \emph{IEEE Global Conf. Signal and
  Information Processing (GlobalSIP)}, Nov. 2017, p. 1080–1084.

\bibitem{cavraro2018GraphAlgorithmToplogyIndenfication}
G.~Cavraro and V.~Kekatos, ``Graph algorithms for topology identification using
  power grid probing,'' \emph{IEEE Control Systems Letters}, vol.~2, no.~4, p.
  689–694, Oct. 2018.

\bibitem{grotas2018PowerSystemsTopology}
S.~{Grotas}, Y.~{Yakoby}, I.~{Gera}, and T.~{Routtenberg}, ``Power systems
  topology and state estimation by graph blind source separation,'' \emph{IEEE
  Trans. Signal Process.}, vol.~67, no.~8, p. 2036–2051, April 2019.

\bibitem{vukovic2012NetworkAwareMitigation}
O.~Vukovic, K.~C. Sou, G.~Dan, and H.~Sandberg, ``Network-aware mitigation of
  data integrity attacks on power system state estimation,'' \emph{IEEE Journal
  on Selected Areas in Communications}, vol.~30, no.~6, p. 1108–1118, Jul.
  2012.

\bibitem{wang2017EffectsSwitchingNetworkTopologies}
S.~Wang, W.~Ren, and U.~M. Al-Saggaf, ``Effects of switching network topologies
  on stealthy false data injection attacks against state estimation in power
  networks,'' \emph{IEEE Syst. J.}, vol.~11, no.~4, p. 2640–2651, Dec. 2017.

\bibitem{kosut2011MaliciousDataAttacks}
O.~Kosut, L.~Jia, R.~J. Thomas, and L.~Tong, ``Malicious data attacks on the
  smart grid,'' \emph{IEEE Trans. Smart Grid}, vol.~2, no.~4, p. 645–658,
  Dec. 2011.

\bibitem{huang2016RealTimeDetection}
Y.~Huang, J.~Tang, Y.~Cheng, H.~Li, K.~A. Campbell, and Z.~Han, ``Real-time
  detection of false data injection in smart grid networks: An adaptive {CUSUM}
  method and analysis,'' \emph{IEEE Syst. J.}, vol.~10, no.~2, p. 532–543,
  Jun. 2016.

\bibitem{jiang2017DefenseMechanisms}
J.~Jiang and Y.~Qian, ``Defense mechanisms against data injection attacks in
  smart grid networks,'' \emph{IEEE Commun. Mag.}, vol.~55, no.~10, p. 76–82,
  Oct. 2017.

\bibitem{ashok2018OnlineDetectionStealthy}
A.~Ashok, M.~Govindarasu, and V.~Ajjarapu, ``Online detection of stealthy false
  data injection attacks in power system state estimation,'' \emph{IEEE Trans.
  Smart Grid}, vol.~9, no.~3, p. 1636–1646, May 2018.

\bibitem{liu2014DetectingFalseData}
L.~Liu, M.~Esmalifalak, Q.~Ding, V.~A. Emesih, and Z.~Han, ``Detecting false
  data injection attacks on power grid by sparse optimization,'' \emph{IEEE
  Trans. Smart Grid}, vol.~5, no.~2, p. 612–621, Mar. 2014.

\bibitem{esmalifalak2017DetectionStealthy}
M.~Esmalifalak, L.~Liu, N.~Nguyen, R.~Zheng, and Z.~Han, ``Detecting stealthy
  false data injection using machine learning in smart grid,'' \emph{IEEE Syst.
  J.}, vol.~11, no.~3, p. 1644–1652, Sep. 2017.

\bibitem{xu2018AchievingEfficientDetection}
R.~Xu, R.~Wang, Z.~Guan, L.~Wu, J.~Wu, and X.~Du, ``Achieving efficient
  detection against false data injection attacks in smart grid,'' \emph{IEEE
  Access}, vol.~5, p. 13787–13798, 2017.

\bibitem{leirzapf2017NetworkForensicAnalysis}
M.~Leierzapf and J.~Rrushi, ``Network forensic analysis of electrical
  substations automation traffic,'' in \emph{Critical Infrastructure Protection
  XI}.\hskip 1em plus 0.5em minus 0.4em\relax Springer, 2017.

\bibitem{yu2015FalseDataPCA}
Z.~H. Yu and W.~L. Chin, ``Blind false data injection attack using {PCA}
  approximation method in smart grid,'' \emph{IEEE Trans. Smart Grid}, vol.~6,
  no.~3, p. 1219–1226, May 2015.

\bibitem{chaojun2015DetectinFalseData}
G.~Chaojun, P.~Jirutitijaroen, and M.~Motani, ``Detecting false data injection
  attacks in ac state estimation,'' \emph{IEEE Trans. Smart Grid}, vol.~6,
  no.~5, p. 2476–2483, Sep. 2015.

\bibitem{hug2012VulnerabilityAssessmentAC}
G.~Hug and J.~A. Giampapa, ``Vulnerability assessment of ac state estimation
  with respect to false data injection cyber-attacks,'' \emph{IEEE Trans. Smart
  Grid}, vol.~3, no.~3, p. 1362–1370, Sep. 2012.

\bibitem{liu2017FalseDataAttackAC}
X.~Liu and Z.~Li, ``False data attacks against {AC} state estimation with
  incomplete network information,'' \emph{IEEE Trans. Smart Grid}, vol.~8,
  no.~5, p. 2239–2248, Sep. 2017.

\bibitem{shuman2013EmergingFieldGSP}
D.~I. Shuman, S.~K. Narang, P.~Frossard, A.~Ortega, and P.~Vandergheynst, ``The
  emerging field of signal processing on graphs: Extending high-dimensional
  data analysis to networks and other irregular domains,'' \emph{IEEE Signal
  Process. Mag.}, vol.~30, no.~3, p. 83–98, May 2013.

\bibitem{sandryhaila2014DSPonGraph}
A.~Sandryhaila and J.~M.~F. Moura, ``Discrete signal processing on graphs:
  Frequency analysis,'' \emph{IEEE Trans. Signal Process.}, vol.~62, no.~12, p.
  3042–3054, Jun. 2014.

\bibitem{teke2017ExtendingClassicalMultirate}
O.~Teke and P.~P. Vaidyanathan, ``Extending classical multirate signal
  processing theory to graphs — {Part I}: Fundamentals,'' \emph{IEEE Trans.
  Signal Process.}, vol.~65, no.~2, p. 409–422, Jan. 2017.

\bibitem{marques2017StationaryGraphProcesses}
A.~G. Marques, S.~Segarra, G.~Leus, and A.~Ribeiro, ``Stationary graph
  processes and spectral estimation,'' \emph{IEEE Trans. Signal Process.},
  vol.~65, no.~22, p. 5911–5926, Nov. 2017.

\bibitem{ortega2018GraphSignalProcessing}
A.~Ortega, P.~Frossard, J.~Kovačević, J.~M.~F. Moura, and P.~Vandergheynst,
  ``Graph signal processing: Overview, challenges, and applications,''
  \emph{Proc. IEEE}, vol. 106, no.~5, p. 808–828, May 2018.

\bibitem{huang2018GSPBrain}
W.~Huang, T.~A.~W. Bolton, J.~D. Medaglia, D.~S. Bassett, A.~Ribeiro, and
  D.~V.~D. Ville, ``A graph signal processing perspective on functional brain
  imaging,'' \emph{Proc. IEEE}, vol. 106, no.~5, p. 868–885, May 2018.

\bibitem{he2018NonIntrusiveLoadDisaggregation}
K.~He, L.~Stankovic, J.~Liao, and V.~Stankovic, ``Non-intrusive load
  disaggregation using graph signal processing,'' \emph{IEEE Trans. Smart
  Grid}, vol.~9, no.~3, p. 1739–1747, May 2018.

\bibitem{drayer2018BadDataGSP}
E.~{Drayer} and T.~{Routtenberg}, ``Detection of false data injection attacks
  in power systems with graph fourier transform,'' in \emph{2018 IEEE Global
  Conf. Signal and Information Processing (GlobalSIP)}, Nov. 2018, p.
  890–894.

\bibitem{wood1984PowerGenerationOperation}
A.~J. Wood and B.~F. Wollenberg, \emph{Power generation, operation, and
  control.}\hskip 1em plus 0.5em minus 0.4em\relax New York : Wiley, C1984,
  1984.

\bibitem{scheme14bus2019}
{University of Washington, Rich Christine}, ``14 bus power flow test case,''
  \url{https://labs.ece.uw.edu/pstca/pf14/pg_tca14bus.htm}, {Accesed:
  20/03/2019}.

\bibitem{rudin1987RealComplexAnalysis}
W.~Rudin and P.~R. Devine, \emph{Real and complex analysis: mathematics series;
  3rd ed.}, ser. Mathematics Series.\hskip 1em plus 0.5em minus 0.4em\relax New
  York, NY: McGraw-Hill, 1987.

\bibitem{hao2015SparseMaliciousFDI}
J.~Hao, R.~J. Piechocki, D.~Kaleshi, W.~H. Chin, and Z.~Fan, ``Sparse malicious
  false data injection attacks and defense mechanisms in smart grids,''
  \emph{IEEE Trans. Ind. Informat.}, vol.~11, no.~5, p. 1–12, Oct. 2015.

\bibitem{zhu2012ApproximatingSignalsOnGraphs}
X.~Zhu and M.~Rabbat, ``Approximating signals supported on graphs,'' in
  \emph{2012 IEEE Int. Conf. Acoustics, Speech and Signal Processing (ICASSP)},
  Mar. 2012, p. 3921–3924.

\bibitem{sandryhaila2013DiscreteSignalProcessingGraphs}
A.~Sandryhaila and J.~M.~F. Moura, ``Discrete signal processing on graphs,''
  \emph{IEEE Trans. Signal Process.}, vol.~61, no.~7, p. 1644–1656, Apr.
  2013.

\bibitem{silva2016PlantWideFaultDetection}
D.~R.~C. Silva and A.~Ortega, ``Plant-wide fault detection using graph signal
  processing,'' in \emph{2016 IEEE Sensor Array and Multichannel Signal
  Processing Workshop (SAM)}, Jul. 2016.

\bibitem{zhu2012GraphSpectralCompressedSensing}
X.~Zhu and M.~Rabbat, ``Graph spectral compressed sensing for sensor
  networks,'' in \emph{2012 IEEE Int. Conf. Acoustics, Speech and Signal
  Processing (ICASSP)}, Mar. 2012, p. 2865–2868.

\bibitem{thurner2018Pandapower}
L.~Thurner, A.~Scheidler, F.~Schafer, J.~H. Menke, J.~Dollichon, F.~Meier,
  S.~Meinecke, and M.~Braun, ``Pandapower - {An} open source python tool for
  convenient modeling, analysis and optimization of electric power systems,''
  \emph{IEEE Trans. Power Syst.}, vol.~33, no.~6, p. 6510–6521, 2018.

\end{thebibliography}


\begin{IEEEbiography}[{\includegraphics[width=1in,height=1.25in,clip,keepaspectratio]{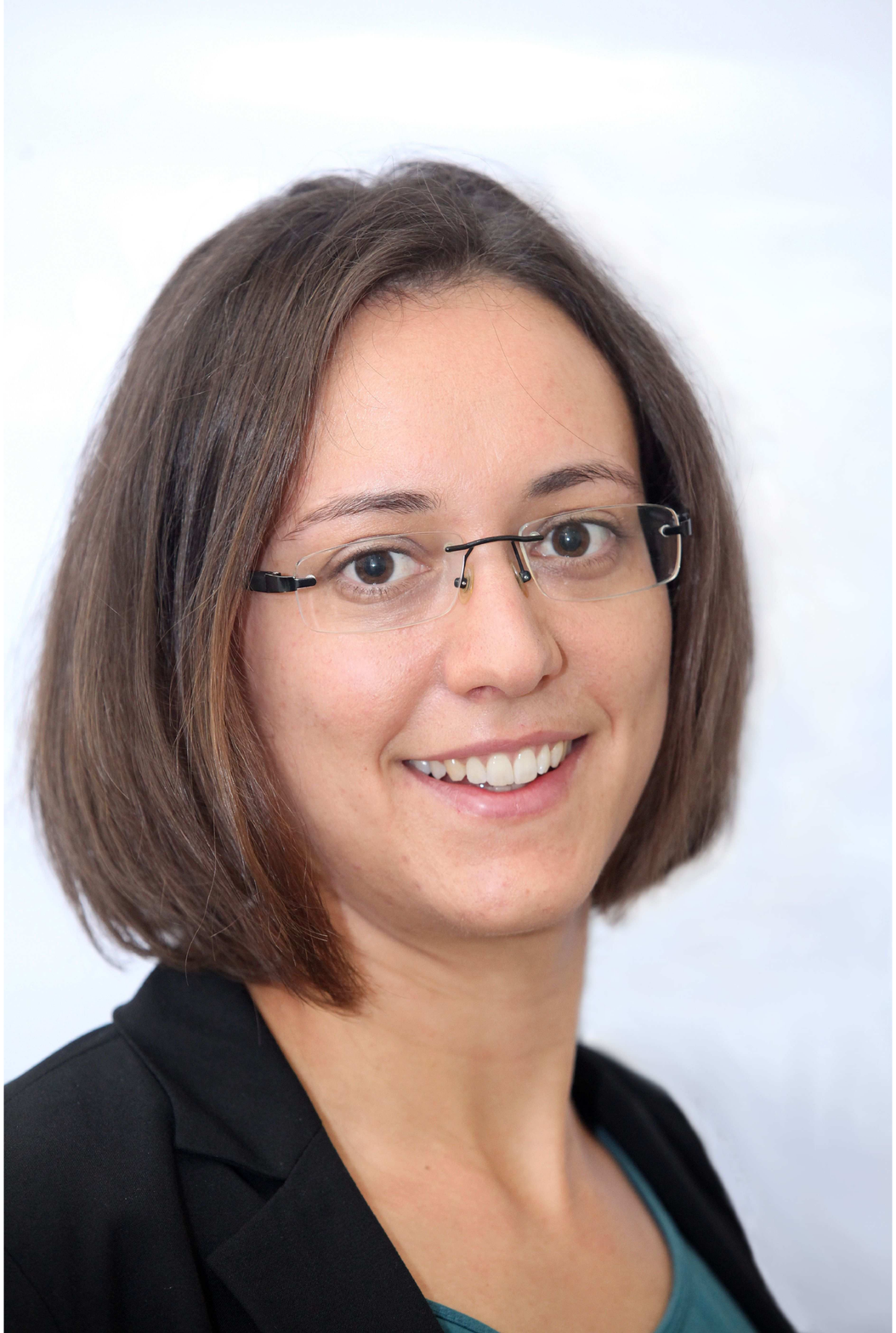}}]{Elisabeth~Drayer}
received a diploma in physics from the Karlsruhe Institute of Technology, Karlsruhe, Germany and a Master in “Energie électrique” from Grenoble INP, Grenoble, France. In 2018 she received her Ph. D. degree in electrical engineering and computer science from the University of Kassel, Kassel, Germany. She is currently a Postdoctoral Fellow at the Ben-Gurion University of the Negev, Beer Sheva, Israel. Her research interests lie in the field of resilient and secure smart grids especially with regard to cyber threats and the application of mathematical concepts to optimize the operation of power systems. 
\end{IEEEbiography}

\begin{IEEEbiography}[{\includegraphics[width=1in,height=1.25in,clip,keepaspectratio]{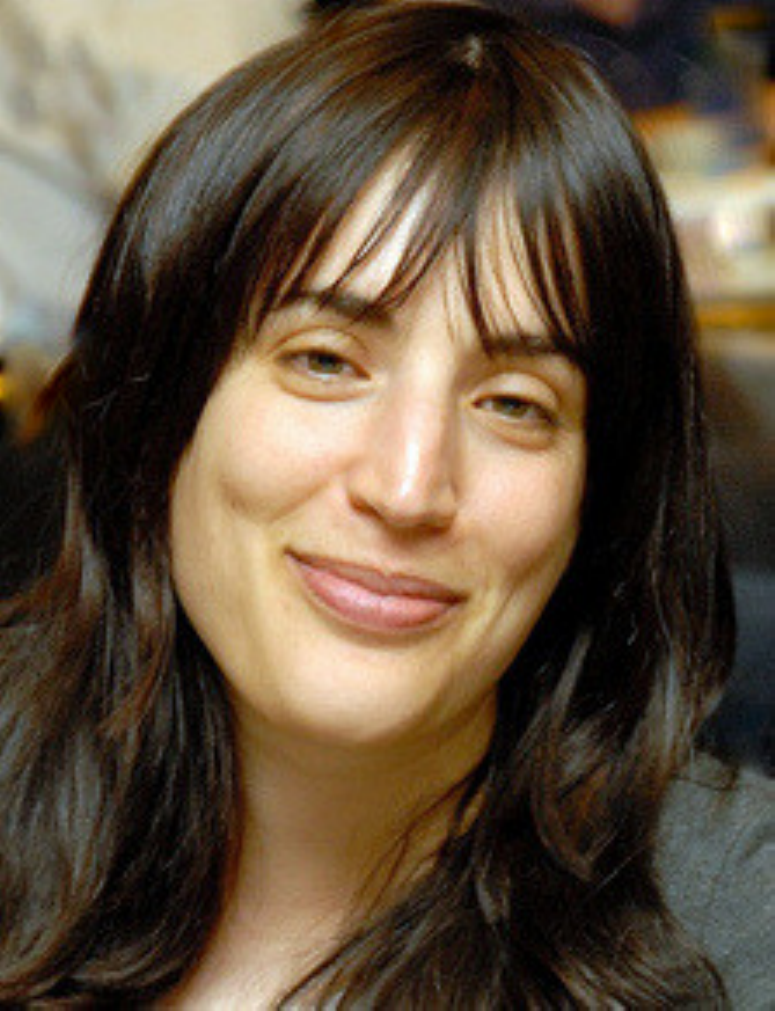}}]{Tirza~Routtenberg}
(S'07-M'13-SM'18) received the B.Sc. degree (magna cum laude) in biomedical engineering from the Technion Israel Institute of Technology, Haifa, Israel, in 2005, and the M.Sc. (magna cum laude) and Ph.D. degrees in electrical engineering from the Ben-Gurion University of the Negev, Beer Sheva, Israel, in 2007 and 2012, respectively. She was a Postdoctoral Fellow with the School of Electrical and Computer Engineering, Cornell University, in 2012-2014. Since October 2014, she is a faculty member at the Department of Electrical and Computer Engineering, Ben-Gurion University of the Negev, Beer Sheva, Israel. Her research interests include signal processing in smart grid, statistical signal processing, estimation and detection theory, and graph signal processing. She was a recipient of the Best Student Paper Award in International Conference on Acoustics, Speech and Signal Processing (ICASSP) 2011, in IEEE International Workshop on Computational Advances in Multi-Sensor Adaptive Processing (CAMSAP) 2013 (coauthor), in ICASSP 2017 (coauthor), and in IEEE Workshop on Statistical Signal Processing (SSP) 2018 (coauthor). She was awarded the Negev scholarship in 2008, the Lev-Zion scholarship in 2010, and the Marc Rich foundation prize in 2011.
\end{IEEEbiography}

%
%
%
%
%




\end{document}